\documentclass[twocolumn,showpacs,amsmath,amssymb,prb,footinbib,floatfix,superscriptaddress]{revtex4-1}

\usepackage{graphicx}
\usepackage{dcolumn}
\usepackage{bm}
\usepackage{helvet}
\usepackage{amssymb}
\usepackage{amsmath}
\usepackage{amsfonts}
\usepackage{hyperref}
\usepackage{color}
\newcommand{\bra}[1]{\langle#1|}
\newcommand{\ket}[1]{|#1\rangle}
\newcommand{\beq}{\begin{equation}}
\newcommand{\eeq}{\end{equation}}

\newcommand{\sandwich}[3]{\langle#1|#2|#3\rangle}

\date{\today}
\begin{document}

\title{Dynamics of entanglement entropy and entanglement spectrum crossing a quantum phase transition}

\author{Elena Canovi}
\affiliation{Institut f\"ur Theoretische Physik III, Universit\"at Stuttgart, Pfaffenwaldring 57, 70550 Stuttgart, Germany}

\author{Elisa Ercolessi}
\affiliation{Dipartimento di Fisica e Astronomia dell'Universit\`a di Bologna and INFN, Sezione di Bologna, Via Irnerio 46, 40127 Bologna, Italy}

\author{Piero Naldesi}
\affiliation{Dipartimento di Fisica e Astronomia dell'Universit\`a di Bologna and INFN, Sezione di Bologna, Via Irnerio 46, 40127 Bologna, Italy}

\author{Luca Taddia}
\affiliation{Dipartimento di Fisica e Astronomia dell'Universit\`a di Bologna and INFN, Sezione di Bologna, Via Irnerio 46, 40127 Bologna, Italy}

\author{Davide Vodola}
\affiliation{Dipartimento di Fisica e Astronomia dell'Universit\`a di Bologna and INFN, Sezione di Bologna, Via Irnerio 46, 40127 Bologna, Italy}
\affiliation{IPCMS (UMR 7504) and ISIS (UMR 7006), Universit\'e de Strasbourg and CNRS, Strasbourg, France}

\begin{abstract}
We study the time evolution of entanglement entropy and entanglement spectrum in a finite-size system which crosses a quantum phase transition at different speeds. We focus on the transverse-field Ising model with a time-dependent magnetic field, which is linearly tuned on a time scale $\tau$.  The time evolution of the  entanglement entropy displays different regimes depending on the value of $\tau$, showing also oscillations which depend on the instantaneous energy spectrum. The entanglement spectrum is characterized  by a rich dynamics where multiple crossings take place with a gap-dependent frequency. Moreover, we investigate the Kibble-Zurek scaling of entanglement entropy and Schmidt gap.
\end{abstract}

\pacs{05.30.Rt, 64.70.Tg, 03.67.Mn, 05.70.Jk}

\maketitle
\section{Introduction}

In recent times, there have been considerable experimental and consequent theoretical advances in the study of the dynamics of closed quantum many-body systems 
(for a review of both the experimental and theoretical aspects, see Ref.~\onlinecite{PSSV2011}). In this work we will deal with the problem of studying the time 
evolution of a closed quantum many body system at $T=0$, when it is driven from one phase to another by allowing the coupling constants in the Hamiltonian to 
change in time. Typical evolution protocols are the so-called slow quenches, where the velocities of the variations of the couplings are finite, 
or, on the contrary, sudden quenches, when such couplings are instantaneously varied to a different value and then left constant. Here we will discuss the 
physical behaviour of an integrable system, namely the Ising chain in a transverse field\cite{LiebSchultzMattis1961,Sachdev:book,Franchini2011}, in the whole range of velocities according to which we let the magnetic field to vary. 

One-dimensional problems may be tackled with generally powerful numerical methods such as t-DMRG\cite{DKSV2004,WhiteFeiguin2004,FeiguinWhite2005} or TEBD\cite{Vidal2004}. It is well known that the final efficiency of such methods is related to the amount of entanglement of the considered state\cite{CiracVerstraete2009}, a quantity which is expected to diverge when getting closer to a phase transition. However, at least in the static case, the behaviour of entanglement  (and more specifically of entanglement entropy) has an universal character so that it can be used as an estimator of quantum correlations\cite{AFOV2008} and to detect as well as to classify  quantum phase transitions also in fully interacting models\cite{XavierAlcaraz2011,Nishimoto2011,DalmonteErcolessiTaddia2011,DalmonteErcolessiTaddia2012,Ercolessietal,Citroetal}. 

Thus, it is natural to ask whether the dynamical behaviour of a closed quantum system, especially when crossing a phase transition, can be described by looking at the dynamics of entanglement entropy and entanglement spectrum, a topic on which there are only a few general results\cite{CalabreseCardy2005,Cincio_PRA07,CanevaFazioSantoro2008,Pollmann_PRE10}. 

The aim of this work is to investigate this question in a paradigmatic example: the Ising chain in a  time-dependent transverse field, a problem which allows  for an exact solution at any instant of time\cite{LiebSchultzMattis1961,Sachdev:book,Franchini2011}. The plan of the work is the following.
In Sec.~\ref{II} we define the notion of entanglement entropy and entanglement spectrum and present the model and its phases. In Sec.~\ref{pf} we describe the dynamics when letting the system go from the paramagnetic to the ferromagnetic phase by controlling the speed with which we change the magnetic field. We will examine the adiabatic regime, the sudden-quench situation and the cases with intermediate speeds. Then we will see how these results are related to the so-called Kibble-Zurek mechanism\cite{Kibble,Zurek} in its quantum version\cite{ZurekDornerZoller2005}, by looking both at the scaling of entanglement entropy and the so-called Schmidt gap\cite{Dechiara_PRL12, GMDADSI2013} in the entanglement spectrum. In Sec.~\ref{IV} we perform a similar analysis when the system evolves from the ferromagnetic to the paramagnetic phase. We end the work with conclusions and outlooks in Sec.~\ref{conc}, and with three Appendices where we have reported technical details of the calculations.

\section{The model}\label{II}

In this work, we are interested in the time evolution of bipartite quantities, such as the entanglement entropy and the entanglement spectrum, 
which are defined in the following way~\cite{NielsenChuang2000}. Starting from a one-dimensional chain of $L$ sites, 
we consider a subsystem $A$  containing $\ell<L$ adjacent sites, $\bar{A}$ being its complement. 
The reduced density matrix is obtained from the pure density matrix of the ground state of the whole chain, $\rho=\left|GS\right>\left<GS\right|$, as
\begin{equation}
 \rho_A=\mbox{Tr}_{\bar{A}}\rho
\end{equation}
The so called entanglement spectrum is the set $\{\lambda_n\}$ of the eigenvalues of $\rho_A$; the entanglement entropy is defined as
\begin{equation}
 S=-\mbox{Tr}_A\rho_A\log_2\rho_A
\end{equation}
and computed as
\begin{equation}
 S=-\sum_n\lambda_n\log_2\lambda_n
\end{equation}
In what follows we take as subsystem the half-chain, i.e. $\ell=L/2$. 
We checked that the main findings of this work are not changed qualitatively if we take a different $\ell$.

In the following, we consider the Ising model in a transverse field\cite{LiebSchultzMattis1961,Sachdev:book,Franchini2011}:
\beq\label{eq:Ising}
 H=-\frac{J}{2}\sum_{j=1}^L\left[\sigma_j^z\sigma_{j+1}^z+h\sigma_j^x\right]
\eeq
where $L$ is the system size and we have periodic boundary conditions (PBC); $\vec{\sigma}$ are the Pauli matrices, and $h=h(t)$ is a time-dependent magnetic field (moreover, we assume $\hbar=J=1$, so that the energies are measured in units of $J$). As recalled in Appendix~\ref{solution}, the model is exactly solvable by a sequence of Jordan-Wigner - Fourier - Bogolyubov transformations; the eigenenergies and the corresponding eigenstates are completely known. Remarkably, the spectrum divides into two sectors, labelled by the quantum number $\alpha\equiv\prod_{j=1}^L\sigma_j^x=\pm1$; at finite size, the ground state always belongs to the $\alpha=1$ sector~\cite{CalabreseEsslerFagotti2012}, which is the one we will deal with in this work. Moreover, the model is one of the prototypical playgrounds for quantum phase transitions~\cite{Sachdev:book}.
Indeed, varying $h$, the Hamiltonian in Eq.~\ref{eq:Ising} displays (in the thermodynamic limit) a quantum critical point at $h=1$, which separates the {\it paramagnetic} ($h>1$) and the {\it ferromagnetic} ($0\leq h<1$) phases~\cite{Sachdev:book} (the negative part of the phase line is the mirror-reflected of the positive one, because of the $\mathbb{Z}_2$ symmetry under the canonical transformation $\sigma_j^x\rightarrow-\sigma_j^x$~\cite{Franchini2011}).
The low-energy physics of such a quantum critical point is described by a conformal field theory\cite{DiFrancescoMathieuSenechal1997} of central charge $c=1/2$. Its correlation-length and dynamic critical exponents are given by $\nu=z=1$~\cite{MorandiNapoliErcolessi2001}.

We make the Hamiltonian in Eq.~\ref{eq:Ising} explicitly time-dependent by letting $h=h(t)$ change linearly in time, from an initial value $h_{i}$ to a final one $h_{f}$:
\beq\label{eq:ramping}
h(t)=h_{i}+{\rm sgn}(h_{f}-h_{i}) \frac{t}{\tau}
\eeq
where $\tau$ is the time scale of the ramping and $t\in [0,t_{f}]$, with $t_{f}=|h_{f}-h_{i}|\tau$ (time is measured in units of $1/J$). The dynamics of the model is also exactly accessible~\cite{McCoy_PRA71}, as we recall in Appendix~\ref{dynamics} (see also Ref.~\onlinecite{PerkAuYang2009}). As we shall see, very different dynamic behaviors are expected for different values of $\tau$.
%
%

%
%

\section{Paramagnet to ferromagnet}\label{pf}
In this Section  we study the ramping from the paramagnetic sector of the phase diagram ($h_i>1$)  to the ferromagnetic one  ($h_f<1$) . This is  the setting for the study of the Kibble-Zurek mechanism in the one-dimensional quantum Ising model\cite{ZurekDornerZoller2005}.
\subsection{Initial structure of the entanglement spectrum}\label{static_ES}
Let us start by studying the initial condition for bipartite quantities. To this aim it is useful to first understand the limit $h_i\rightarrow\infty$.
 In this case
 the ground state of the system at $t=0$ looks like
 \begin{equation}
  \left|0\right>\equiv\prod_{j=1}^L\left|\rightarrow\right>_j
 \end{equation}
where we denote with $\left|\rightarrow\right>_j,~\left|\leftarrow\right>_j$ the state with $S_j^x= \pm \hbar/2$ respectively.
 Of course, this is not the exact ground state for finite $h_i\gg1$, but, at first order in perturbation theory, it is easy to show that the latter is given by
 \begin{equation}
  \left|GS\right>=N\left[\left|0\right>+\frac{1}{4h}\sum_{j=1}^L\left|j,j+1\right>\right]\label{para_GS}
 \end{equation}
 with $\left|j,j+1\right>$ being the state with two spin flips at sites $j$, $j+1$:
 \begin{equation}
  \left|j,j+1\right>\equiv\left|\leftarrow\right>_j\left|\leftarrow\right>_{j+1}\prod_{\substack{k=1\\ k\neq j,j+1}}^L\left|\rightarrow\right>_k
 \end{equation}
 $N\equiv\left(1+\frac{L}{16h^2}\right)^{-\frac{1}{4}}$ is the normalization coefficient (this usually neglected normalization factor is necessary to obtain a good agreement with numerical results).

 The zero-temperature density matrix of the system is given by
 \begin{equation}
  \rho=\left|GS\right>\left<GS\right|
 \end{equation}
 and the reduced density matrix $\rho_A=\mbox{Tr}_{\bar{A}}\rho$ of the half chain $A=\{1,\cdots,L/2\}$ (we will always choose this bipartition) is seen to take the form
 \begin{equation}
  \rho_A=\left(\begin{array}{llll}
   \left|0\right>_A, & \left|2p\right>_A, & \left|1\right>_A, & \left|L/2\right>_A
  \end{array}\right)\mathbb{R}_A\left(\begin{array}{l}
   _A\left<0\right|\\
   _A\left<2p\right|\\
   _A\left<1\right|\\ 
   _A\left<L/2\right|
  \end{array}\right)
 \end{equation}
  being $\left|0\right>_A$ the paramagnetic state relative to subsystem $A$. Also:
  \begin{equation}
   \begin{split}
    \left|2p\right>_A&\equiv\left(\frac{L}{2}-1\right)^{-\frac{1}{2}}\sum_{j=1}^{\frac{L}{2}-1}\left|j,j+1\right>,\\
    \left|1\right>_A&\equiv\left|\leftarrow\right>_1\prod_{j=2}^{\frac{L}{2}}\left|\rightarrow\right>_j,\\
    \left|L/2\right>_A&\equiv\left|\leftarrow\right>_{L/2}\prod_{j=1}^{\frac{L}{2}-1}\left|\rightarrow\right>_j
   \end{split}
  \end{equation}
  and
  \begin{equation}
   \mathbb{R}_A\equiv N^2\left(\begin{array}{cccc}
    1+\frac{\frac{L}{2}-1}{16h^2} & \frac{\sqrt{\frac{L}{2}-1}}{4h} & 0 & 0 \\
    \frac{\sqrt{\frac{L}{2}-1}}{4h} & \frac{\frac{L}{2}-1}{16h^2} & 0 & 0 \\
    0 & 0 & \frac{1}{16h^2} & 0 \\
    0 & 0 & 0 & \frac{1}{16h^2}
   \end{array}\right)
  \end{equation}
  The form of $\mathbb{R}_A$ shows that $\left|1\right>_A$ and $\left|L/2\right>_A$ are true eigenstates of $\rho_A$; diagonalizing the remaining block it can be seen that, for large enough $h$, the remaining two eigenstates are superpositions of $\left|0\right>_A$ and $\left|2p\right>_A$,
   one in which the paramagnet dominates and the other in which $\left|2p\right>_A$ dominates. A numerical analysis shows that the largest eigenvalue is associated with the first one, while the smallest with the second; the ones associated with the single flipped states $\left|1\right>_A$ and $\left|L/2\right>_A$ are of course degenerate, and occupy the second and third position in magnitude.

\subsection{General dynamical features}

\begin{figure}[t]
  \includegraphics[width=0.49\textwidth]{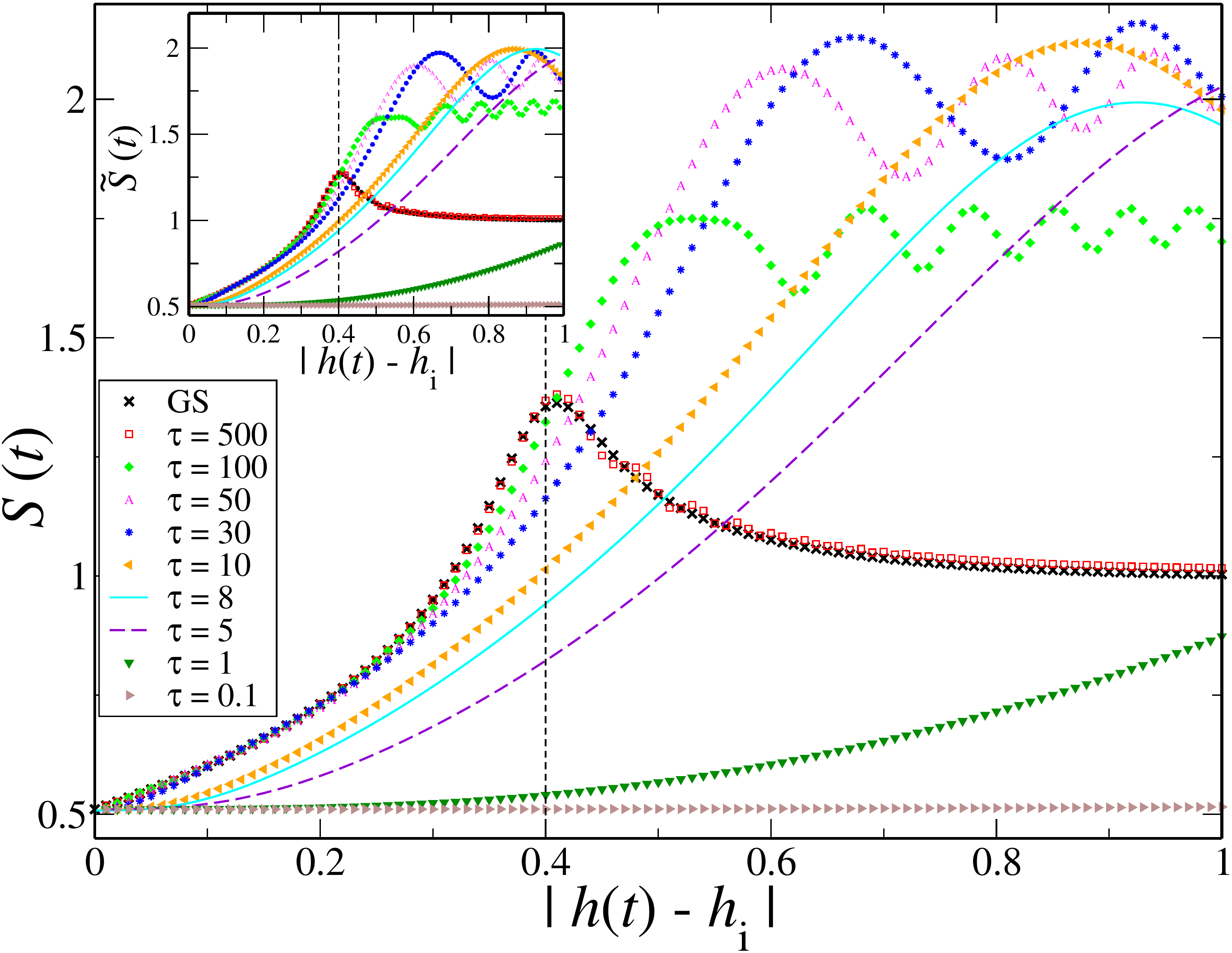}
 \caption{Dynamics of the entanglement entropy for $L=50$, $h_i=1.4$, $h_f=0.4$. Main panel: $S(t)$ for different values of $\tau$ (dashed verical line: location of the critical point). Inset: $\tilde S(t)$ as defined in Eq.~\ref{eq:partent} for the same values of $\tau$ as in the main panel.
 }\label{partial}
\end{figure}

\begin{figure}[t]
  \includegraphics[width=0.45\textwidth]{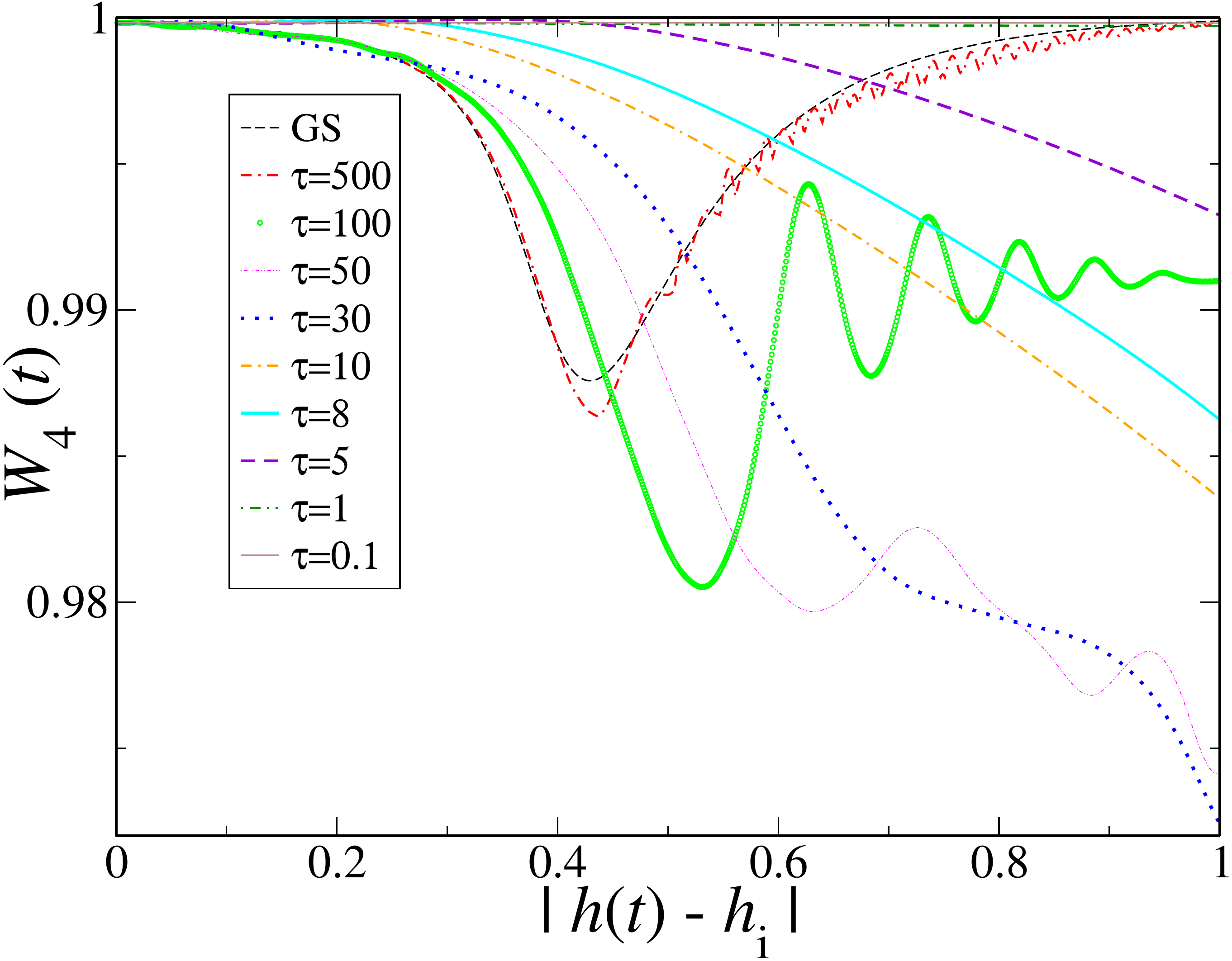}
 \caption{
 Sum of the first four eigenvalues of the reduced density matrix $W_{4}$ (see Eq.~\ref{eq:sum}) for $L=50$, $h_i=1.4$, $h_f=0.4$
 and different values of $\tau$.
 }\label{fig:sum}
\end{figure}

 The aim of this Section is to show that just few eigenvalues contribute to the entanglement dynamics of the system. 
 To see this, we first compute the entanglement entropy for a half-chain bipartition, as explained in Appendix~\ref{bip} (see Eq.~\ref{eq:entropy}).
 The results for $L=50$, $h_i=1.4$, $h_f=0.4$ and different values of $\tau$ are shown in the main panel of Fig.~\ref{partial}. 
 Then, we compute the first four eigenvalues of the reduced density matrix of $A$ and consider their sum
 \beq\label{eq:sum}
 W_{4}(t)=\sum_{n=1}^{4}\lambda_{n}(t)\;,
 \eeq
 which is shown in Fig.~\ref{fig:sum}. 
 We notice that $W_{4}$ is very close to unity for fast rampings ($\tau\lesssim 1$) and, away from the critical value of $h$, 
 for nearly adiabatic rampings ($\tau \gtrsim 500$). In all other cases the weight of the first four eigenvalues 
 is always at least $\sim 0.97$; moreover, this seems to be quite $L$-independent (not shown).  Therefore, in the following, we will always consider just the first four eigenvalues of the reduced density matrix.
 By means of the latter eigenvalues, we can compute a ``partial'' entanglement entropy:
 %
\begin{equation}\label{eq:partent}
 \tilde S(t)=-\sum_{n=1}^{4}\lambda_n(t)\log_2\lambda_n(t)\;.
\end{equation}
We show this quantity in the inset of Fig.~\ref{partial}, where we see that, apart from a small quantitative discrepancy, 
the qualitative behaviour is indeed the same as the one of the true entanglement entropy $S(t)$.

In Sec.~\ref{adsud}-\ref{slow} we will study the dynamics of the entanglement spectrum  in detail (Fig.~\ref{p->f_fig}). The entanglement spectrum at the initial value $h_i$ is very close to the one described in Sec. \ref{static_ES} (large field case). As we shall see, it displays, as well as entanglement entropy, a different dynamical qualitative behaviour depending on the value of $\tau$. Unless explicitly stated,  we choose $L=50$ (postponing the discussion of size-effects to Sec. \ref{KZ}) 
and show our results for a ramping from $h_i=1.4$ to $h_f=0.4$. We choose these values of the initial and final magnetic field in order to restrict the range of integration of the ODE's, Eq. \ref{ODE}. 

\subsection{Adiabatic and sudden regimes}\label{adsud}
 
\begin{figure*}[t]
\includegraphics[width=0.8\textwidth]{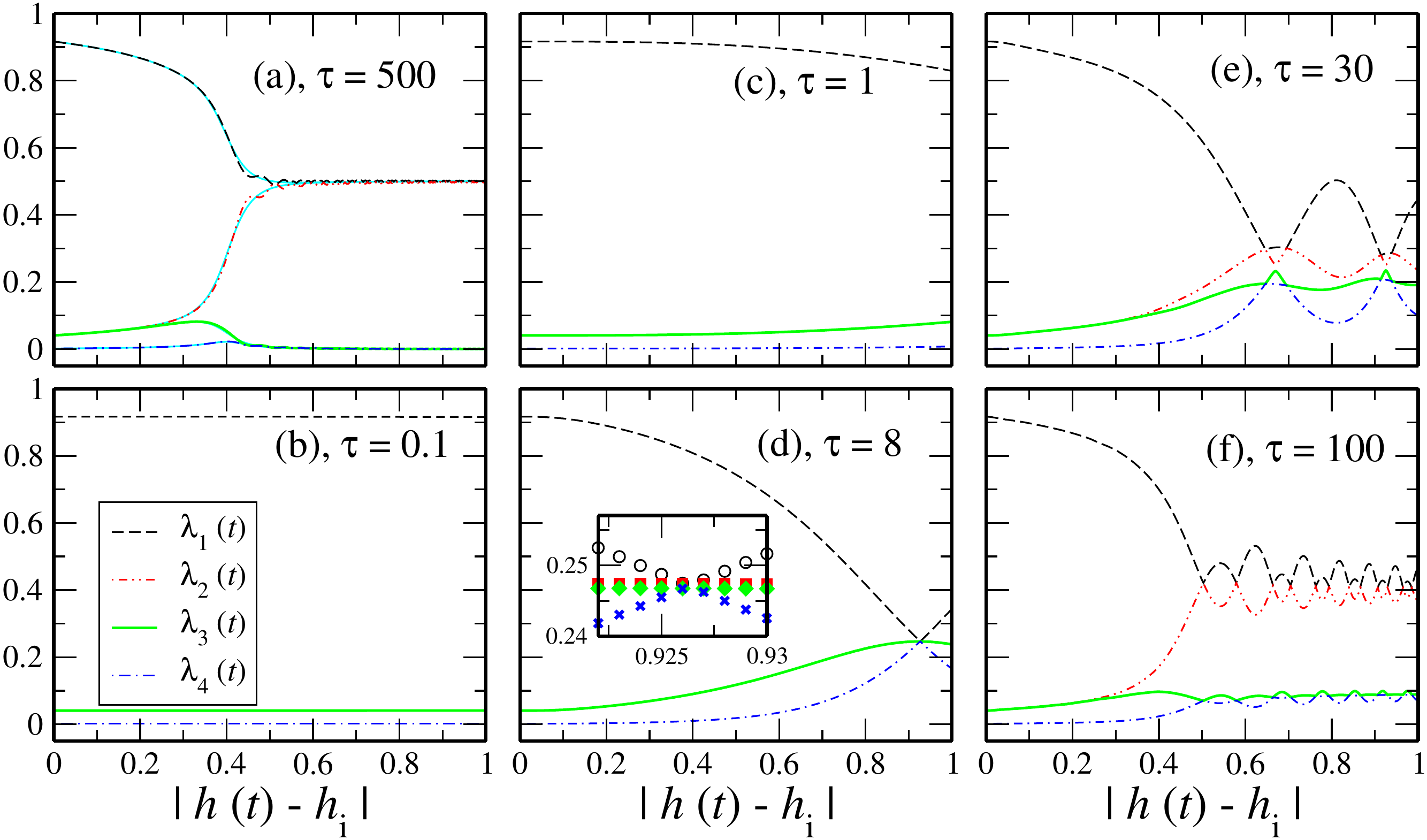}
\caption{Dynamics of the entanglement spectrum for $L=50$, $h_i=1.4$ and $h_f=0.4$. Black dashed, red dot-dot-dashed, green solid and blue dash-dotted lines correspond to the  
 dynamical first, second, third and fourth eigenvalue of the reduced density matrix of the half chain respectively. Different panels refer to (a): $\tau=500$; (b): $\tau=0.1$; (c): $\tau=1$; (d): $\tau=8$; (e): $\tau=30$; (f): $\tau=100$ respectively. In panels (b) , (c) and (d) the red and green lines overlay. The cyan lines in panel (a) show the 
 ground-state values of the first four eigenvalues. The inset in panel (d) is a zoom of the crossing point. }\label{p->f_fig}
\end{figure*}

 We begin by considering very large values of $\tau$, i.e., a quasi-adiabatic quench, see for example the curve at $\tau=500$ of Fig.~\ref{partial} and panel (a) of Fig.~\ref{p->f_fig}.
 We observe that  during the evolution the entanglement entropy and the entanglement spectrum closely follow the {\it static} values, i.e., 
 those obtained from the ground state of the system at each value of $h(t)$, the only difference being
 represented by some small oscillations, that will be discussed in Sec. \ref{slow}.
 This behaviour is expected from the adiabatic theorem\cite{Messiah1999} and as a consequence of the finite size of the system. Indeed the energy gap closes as a function of the inverse size, 
 remaining non-zero for any finite $L$, so that in this case it is always possible to reach the adiabatic limit provided $\tau$ is large enough (see also Sec. \ref{KZ}). More precisely, as shown by Cincio et al. in Ref.\onlinecite{Cincio_PRA07}, the probability of having an adiabatic evolution at size $L$ is given by $P (\tau) = 1-\exp(-2 \pi^3\tau/L^2)$, so that the maximum rate ($\sim1/\tau$) at which the evolution is adiabatic decays as $1/L^2$. 

 We then consider the opposite regime, with very small values of $\tau$, i.e., very fast quenches: we show this situation in Fig.~\ref{partial} (curve with $\tau=0.1$ of the main panel) and Fig. \ref{p->f_fig}(b). 
 The entanglement entropy and the entanglement spectrum do not evolve at all, as expected
 from the adiabatic theorem, independently on the size of the system. 
 %
%

 \subsection{Fast sweeps}\label{fast}
 We consider now rampings that are slower than sudden ones, but much faster than adiabatic ones; 
 we call them {\it fast} sweeps, and, for our system sizes, they correspond to $\tau=1\div20$.  
 For the sake of clarity, for  both the entanglement entropy and the entanglement spectrum 
 it is useful first to consider the faster regime $\tau\sim1$ and then 
 slower rampings $\tau\sim10\div20$. 
 
Starting from faster rampings (see curves with $\tau=1$ and 5 in the main panel of Fig.~\ref{partial}), 
 the entanglement entropy increases linearly in the region close to the phase transition: this behaviour can be related to the results of  Calabrese and Cardy~\cite{CalabreseCardy2005} relative to a sudden quench to a conformal critical point . In their case, the entanglement entropy is predicted to grow (at least in the first part of its evolution) linearly, with a slope related to the central charge of the underlying conformal field theory.
Even if our case is different from the cited one because of the finite ramping speed, we can try to apply this picture. Indeed, close to the critical point, the correlation length and relaxation time are large, so that the system behaves as critical for a finite interval of $h$.

 The behaviour of the entanglement spectrum is of course related to the one of the entanglement entropy and is shown in panel (c) of Fig.~\ref{p->f_fig}. 
 In this regime of $\tau$, the first eigenvalue decreases, while the remaining three increase: this results in a growth of the entanglement entropy that we observe\cite{NielsenChuang2000}. Remarkably, the second and third eigenvalues of the reduced density matrix remain degenerate even during this kind of evolution: indeed, these eigenvalues correspond, at $t=0$, to the eigenstates $\left|1\right>$ and $\left|L/2\right>$ (see Sec. \ref{static_ES}), and the time evolution, as shown by a perturbative analysis (that we are not going to report), does not break this degeneracy, at least for these values of $\tau$.

\begin{figure}[t]
\includegraphics[width=0.49\textwidth]{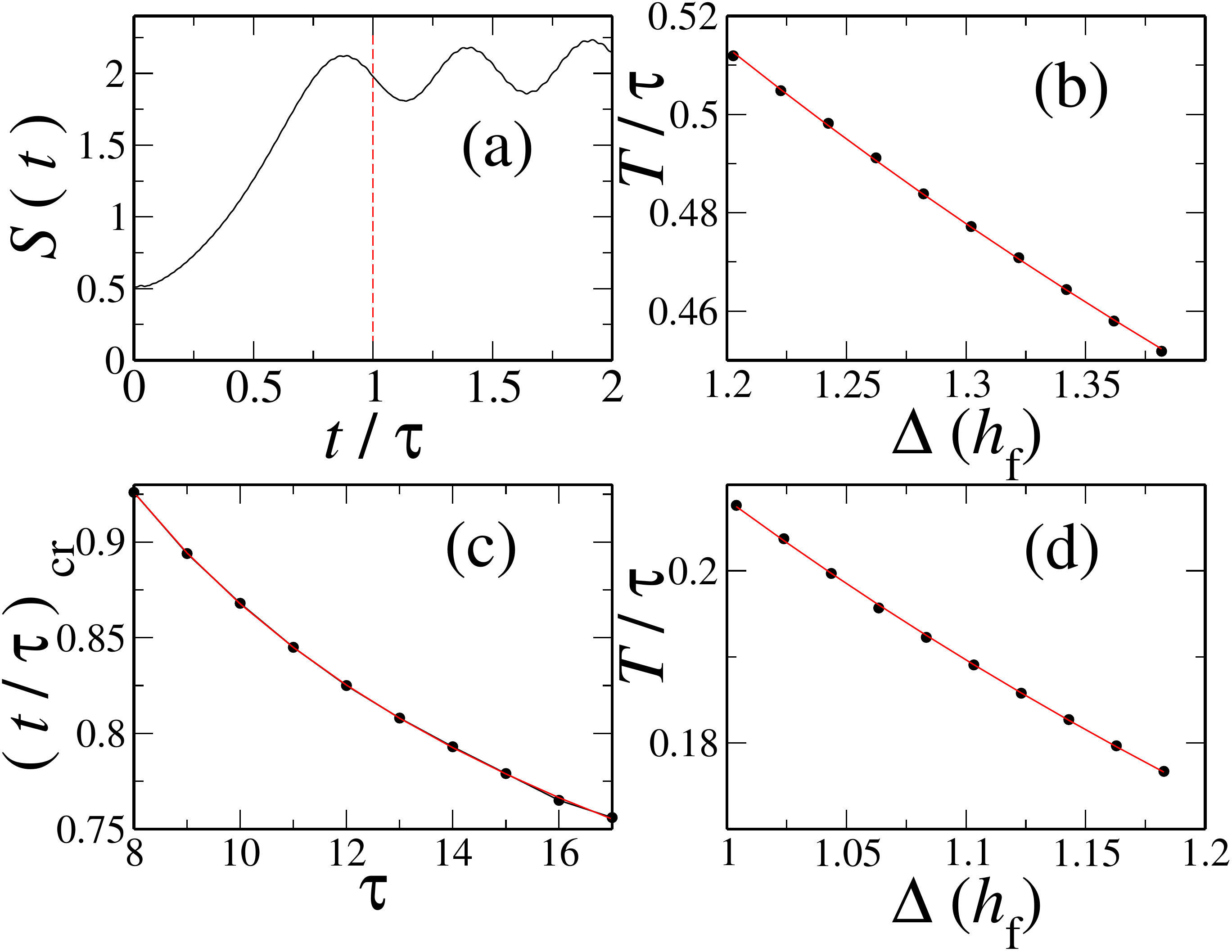}
 \caption{Panel (a): evolution of the entanglement entropy according to the protocol (\ref{waiting}) for $\tau=10$. Panel (b): oscillation period as a function of the energy gap ($h_i=1.4$, $h_f\in[0.31,0.4]$); the fit is performed by means of the formula $y=a_0+a_1/x$. Panel (c): time at which the eigenvalues of the reduced density matrix cross as a function of $\tau$, for $\tau\in[8,17]$. Black dots: numerical data; red line: fitting formula $y=a_0+a_1/x^{a_2}$, giving $a_0=0.428193$ (critical point: 0.4). Panel (d): same as in (b), but with $h_i=1.5$, $h_f\in[0.41,0.5]$ and $\tau=30$. }\label{omega(gap)+crossing(tau)}
 \end{figure}

 The second regime is encountered by further increasing $\tau$ (see for example curves with $\tau=8, 10$ and 30 in the main panel of Fig.~\ref{partial}). In such cases, the entanglement entropy still presents a linear-growth region, which does not last to the end of the sweep, 
 ending in an oscillatory region, in which the entanglement entropy alternates between maxima and minima, with variable frequency. 
 This behaviour has already been observed in a thermodynamic-limit study of the dynamics of entanglement entropy\cite{Pollmann_PRE10}, 
 and has been ascribed to the fact that the system ends up, after passing the critical point, in a superposition of excited states of the instantaneous Hamiltonian.
  In particular, the oscillation frequency has been predicted to scale as
 \begin{equation}
  \omega(t)\approx\Delta(t)
 \end{equation}
 being $\Delta(t)$ the energy gap of the instantaneous Hamiltonian, given by Eq.~\ref{eq:Ising}. 
 To verify this prediction in our finite-size system, we evolve the ground state of $H(t=0)$ according to the protocol
 \begin{equation}\label{waiting}
  h(t)=\left\{\begin{array}{ll}
   h_{i}-\frac{t}{\tau}, & 0\leq t\leq\left(h_i-h_f\right)\tau\\
   h_f, & t>\left(h_i-h_f\right)\tau
  \end{array}\right.
 \end{equation}
 i.e., the usual ramping  of Eq.~\ref{eq:ramping} followed, in the end, by an evolution according to the final Hamiltonian. 
 The result is that in the second part of the evolution the entropy oscillates with a constant frequency 
 (figure \ref{omega(gap)+crossing(tau)}(a)), and such an oscillation is superimposed to another one, much smaller 
 in amplitude and period (plus an increasing power-law trend). To determine the period of the first one, we fit the 
 right part of the curve in figure \ref{omega(gap)+crossing(tau)}(a) by means of the seven-parameters formula 
 $y=a_0+a_1x^{a_2}+a_3\cos\left(\frac{2\pi x}{a_4}+a_5\right)/x^{a_6}$, that turns out to be a very suitable fitting equation 
 (apart from the subdominant oscillatory behaviour); $a_4$ is directly the period of the oscillation. 
 After getting from the fits the values of the periods for several values of $h_f$, we can plot them as a function of $\Delta(t)$ 
 (the gap is computed from the exact solution; see Appendix~\ref{solution}). The results are shown in figure \ref{omega(gap)+crossing(tau)}(b): 
 the behaviour of the oscillations period as a function of the gap is compatible with a $1/x$ law. 
 This check confirms, as we shall see in Sec. \ref{KZ}, that the physics in this regime is the same as in the thermodynamic limit.

 We now investigate the behaviour of the entanglement spectrum in this regime. As shown in Fig.~\ref{p->f_fig}(d), 
 the decreasing of the first eigenvalue and the growth of the remaining ones continues until they cross, all at the same point. 
Moreover, this crossing structure recurs also  for later times in an almost periodic pattern (not shown). This behaviour is very peculiar, and we shall investigate it in detail. 
 First of all, it must be noticed that the crossings correspond, as expected, to  the maxima of entanglement entropy and that this oscillatory behaviour starts only after the system has crossed the critical point.
 This fact is easily confirmed by plotting the crossing time $t_{cr}$ as a function of $\tau$: the result is shown 
 in Fig.~\ref{omega(gap)+crossing(tau)}(c): the data can be fitted by a power-law 
 , showing that, 
 for $\tau\rightarrow\infty$, the crossing point converges with good precision to the critical point (strictly speaking, 
 we could not take the limit $\tau\rightarrow\infty$, since, for larger $\tau$, the behaviour of the system tends 
 to become adiabatic; however, this extrapolation shows that the oscillations, also present for larger $\tau$, 
 always have the same nature; see Sec. \ref{slow}).
We have also verified that the crossing time $t_{cr}$ does not depend on the size of the system at fixed $\tau$ (not shown): this fact represents a further evidence of the fact that the physics, for these values of $\tau$, coincides with the thermodynamic-limit one.

 By magnifying the crossing region (see inset of figure \ref{p->f_fig}(d)), it becomes manifest that the fourfold crossing 
 is actually a crossing between the first and the fourth eigenvalue, while the second and the third continue evolving parallel to each other. 
 
\begin{figure}[t]
\includegraphics[width=0.49\textwidth]{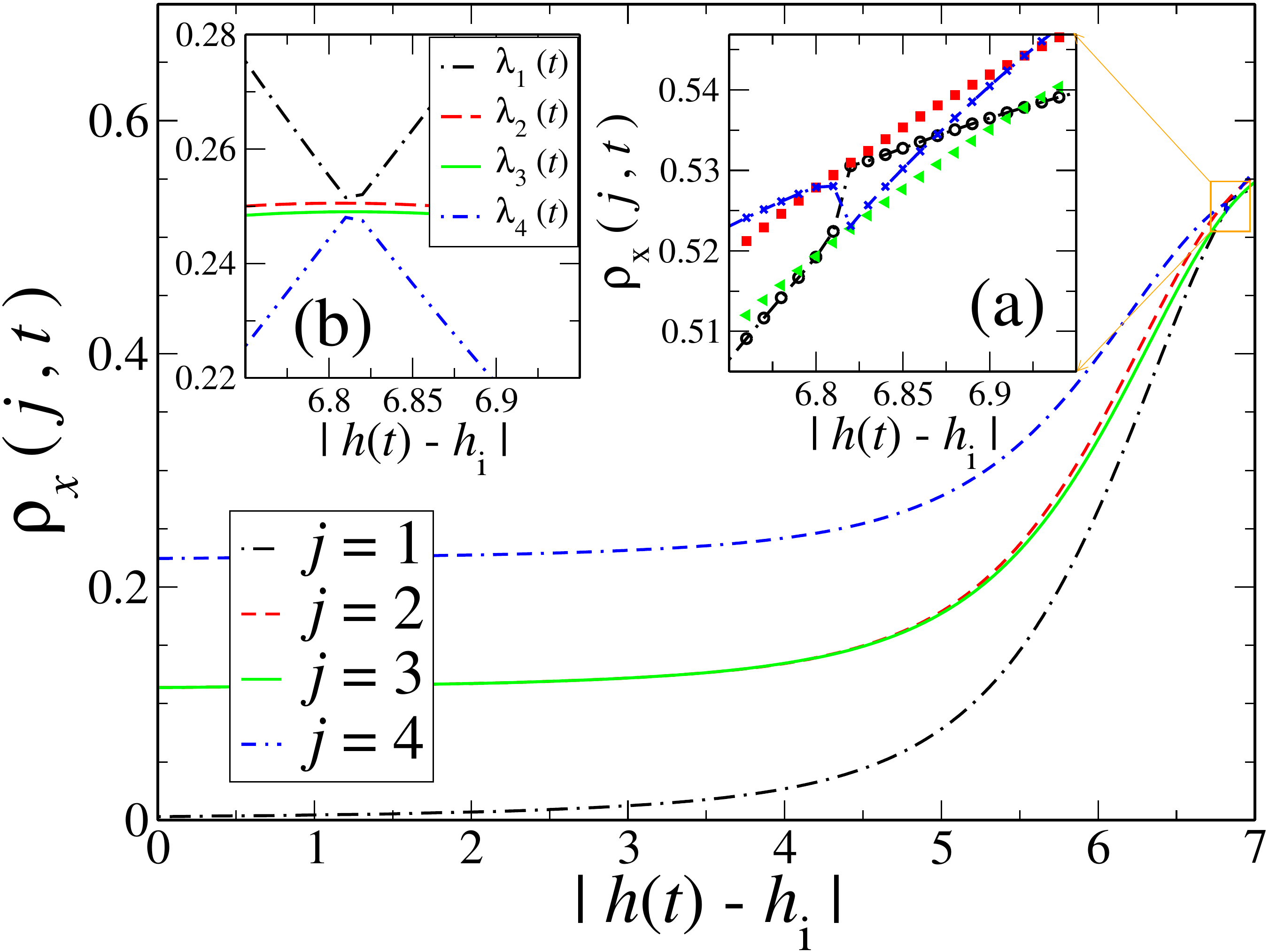}
\caption{Main panel: time evolution of the expectation value of the density of left spins on the first four 
eigenstates of the reduced density matrix (cfr. Eq.~\ref{eq:rhox}) for a ramping of the magnetic field 
from $h_{i}=6$ to $h_{f}=-1$ with $\tau=1$ and $L=18$ sites. Panel (a): zoom of the main panel around the crossing point, with the usual color code. Panel (b): corresponding dynamics of the entanglement spectrum in the same time interval of panel (a).}\label{fig:flips}

\end{figure}

 A question which might arise is what are the signatures of the crossing of the eigenvalues of the reduced density matrix on observable quantities. As an example, we can get an insight in this direction by computing
the expectation value of the density of left spins on each of the eigenstates,
defined by:
%
 \begin{equation}\label{eq:rhox}
  \rho_{x}(j,t)\equiv\frac{1}{\ell}\sum_{i=1}^{\ell} \sandwich{j(t)}{\frac{1}{2}(1-\sigma^{x}_{i})}{j(t)}
 \end{equation}
 being $\left|j(t)\right>$ the $j$-th eigenstate of the reduced density matrix at time $t$ ($j=1,2,3,4$). 
 Since the latter states are many-body objects, in order to compute $\rho_{x}$ we use
 exact diagonalization, with the time evolution performed via time-dependent Lanczos algorithm~\cite{lanczos} (in a recent study\cite{ZRLLT2014}, Zamora and collaborators were able to compute similar quantities in the free-fermions approach).
 Our results are summarized in Fig.~\ref{fig:flips} for a quench with $h_{i}=6$ and $h_{f}=-1$ and $L=18$ sites.
 The choice of this interval of $h$, much larger than the ones considered up to now,
 is due to the small size of the system. However, the crossing of the eigenvalues $\lambda_{i}$ has the same structure as in the inset of Fig. \ref{p->f_fig}(d).
First of all we notice from the main panel that for all the four eigenstates the density $\rho_{x}$ increases 
during the time evolution. This is expected, since the operator $\sigma^{z}_{j}\sigma^{z}_{j+1}$ rules the evolution of the system, by creating
pairs of left spins. 
Almost at the end of the evolution the densities approach each other, in a way which is magnified 
in panel ~\ref{fig:flips}(a). 
The densities corresponding to the first and the fourth eigenvalue, i.e. $\rho_{x}(1,t)$
and $\rho_{x}(4,t)$ are first exchanged at a time $t\sim 6.81$, in correspondence with the crossing of $\lambda_{1}$ and $\lambda_{4}$, and then cross each other a $t\sim 6.86$. The important information which we can extract from Fig. ~\ref{fig:flips}(a) is that the main contribution to the density of flips in the half-chain (which is the physical meaning of $\rho_{x}(1,t)$) changes the profile of its time evolution in correspondence with the crossings of the eigenvalues of the reduced density matrix. In particular, if we follow the black symbols in ~\ref{fig:flips}(a), the density $\rho_{x}(1,t)$ increases more slowly after the crossing.
 \subsection{Slow sweeps}\label{slow}
 The last regime is observed for $\tau\gtrsim20$; a typical situation is shown in Fig.~\ref{p->f_fig}(e), (f). 
 As figure \ref{p->f_fig}(e) shows, the second and the third eigenvalues begin to separate, and then the crossing of the first and the fourth begins to become an avoided crossing; for larger values of $\tau$, as shown in figure \ref{p->f_fig}(f), this separation continues and the dynamical structure of the spectrum gets closer to the static one, i.e., the one of figure \ref{p->f_fig}(a). In all cases, the crossings take now place between the first and second, third and fourth eigenvalues; remarkably, they take place at the same times for the first and the second couple. On the other hand, the entanglement entropy, as shown in the main panel of Fig.~\ref{partial} (curves with $\tau=100$), at the beginning of the evolution is practically coincident with the static one, and at a certain point begins to grow; however, it begins to oscillate around a value that is smaller than the ones of Sec. \ref{fast}, and that decreases as $\tau$ increases. This behaviour of entanglement spectrum and entanglement entropy can 
be ascribed to the approaching of the adiabatic regime. However, as already observed in Sec. \ref{adsud}, the oscillation studied in Sec.~\ref{fast} survive as a sign of non-adiabaticity, but this time between the first and the second two eigenvalues. Even in this case, performing the same analysis as in Sec.~\ref{fast}, we can show that the period of such oscillation at the instant $t$ decreases as a function of the inverse gap of $H(t)$ (see Fig. \ref{omega(gap)+crossing(tau)}(d)).
 
\begin{figure}[t]
\includegraphics[width=0.42\textwidth]{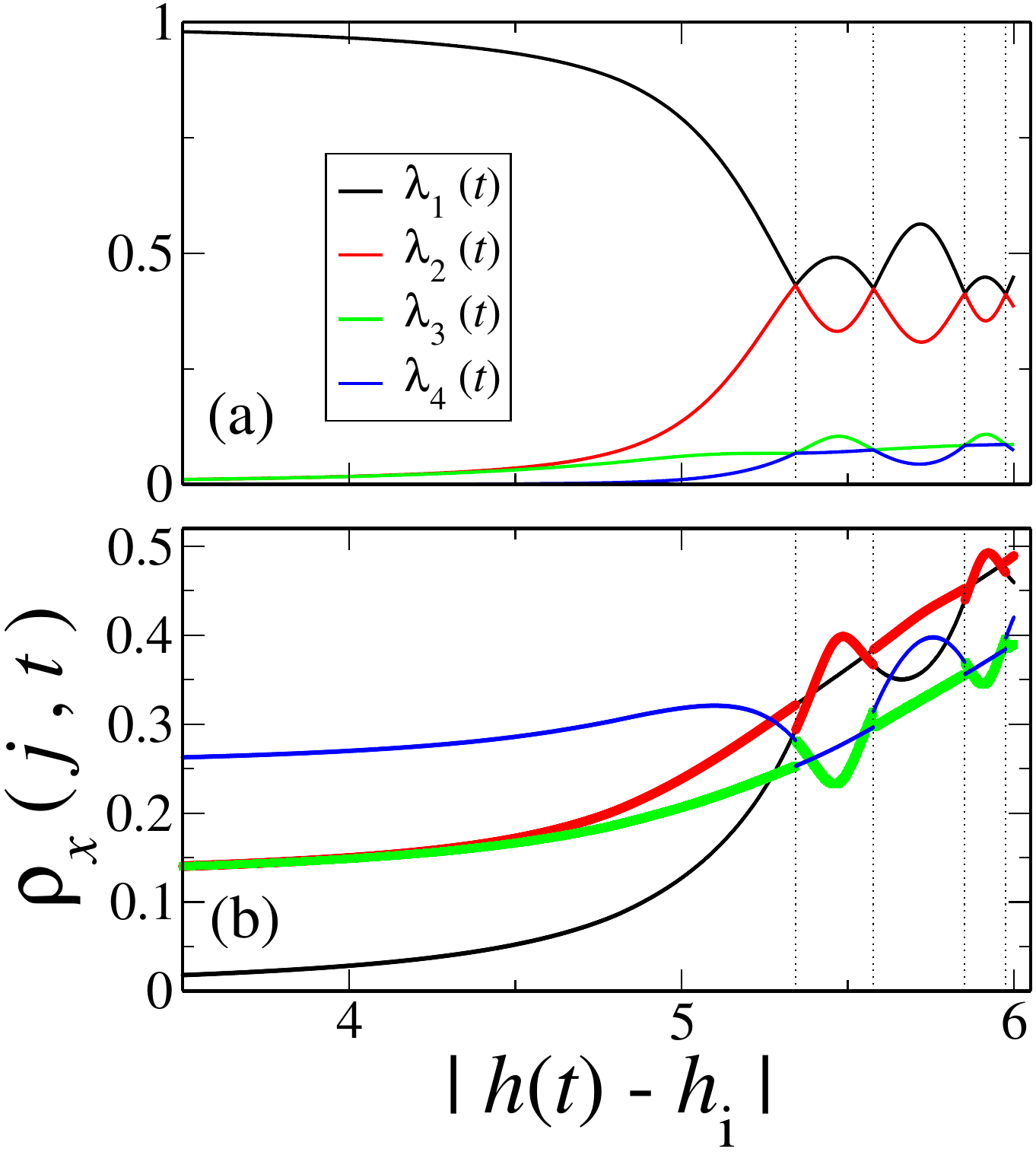}
\caption{Panel (a): dynamics of the entanglement spectrum for a ramping from $h_{i}=6.0$ to $h_{f}=0$, with $\tau=10$ and $L=18$. 
Panel (b): expectation value of the density of left spins. 
}\label{fig:slow}
\end{figure}

%
%
%
By taking a small system with $\tau$ in this slow regime, we can proceed as in Sec.~\ref{fast} to compute the expectation value of the density of left spins (see Eq.~\ref{eq:rhox}); we show our results in Fig.~\ref{fig:slow}. In panel (a) the dynamics of the eigenvalues is shown to be analogous to that of Fig.~\ref{p->f_fig}(f).
In panel (b) of  Fig.~\ref{fig:slow} we first observe a crossing of the density of left spins corresponding to the first and third eigenstate, immediately followed by a crossing of the second and the fourth and another involving the first and the fourth eigenstate. When the crossings of the $\lambda_{j}(t)$ take place, the densities of left spins exchange and, at different times, cross each other, until at the end of the evolution we observe two pairs of self-avoiding levels. Even in this case, the density of left spins is the only quantity displaying crossings in correspondence of the crossing of the eigenvalues (up to a non-synchronization of the exchanges of the eigenvalues and the crossings themselves).


 \subsection{Kibble-Zurek physics}\label{KZ}
 In this Section, we discuss the Kibble-Zurek scaling\cite{Kibble, Zurek} of two quantities, i.e., the already considered entanglement entropy and the Schmidt gap\cite{Dechiara_PRL12,GMDADSI2013}, i.e., the difference between the two largest eigenvalues in the entanglement spectrum.
A discussion of this mechanism for the $XY$-model may be found in Refs.~\onlinecite{ZurekDornerZoller2005}, \onlinecite{Dziarmaga2005}, \onlinecite{Polkovnikov2005} and \onlinecite{MDDS2007}.

 In its original formulation, the Kibble-Zurek mechanism is able, on the basis of extremely simple approximations, to predict the scaling of the number of topological defects produced after the dynamical transition of a critical point. The key assumption underlying the mechanism is that the evolution can be divided, for suitable ramping velocities, into three parts: a first {\it adiabatic} one, where the wave function of the system coincides with the ground state of $H(t)$; a second {\it impulsive}, where the wave function of the system is practically frozen, due to the large relaxation time close to the critical point; a third adiabatic one, as the system is driven away from the critical point~\cite{ZurekDornerZoller2005}. This division takes the name of {\it adiabatic-impulse-adiabatic} approximation\cite{DamskiZurek}. What plays a role in this kind of mechanism is the {\it correlation length} $\hat{\xi}$ at the times of passage between the different regimes, that can be seen to scale, for a linear quench 
of inverse velocity $\tau$, as\cite{Zurek}
 \begin{equation}\label{kz-xi}
  \hat{\xi}\approx\tau^{\frac{\nu}{1+z\nu}}
 \end{equation}
 being $\nu$ and $z$ the critical exponents of the crossed quantum critical point\cite{MorandiNapoliErcolessi2001}.
 \subsubsection{Entanglement entropy}
\begin{figure}[t]
  \includegraphics[width=0.45\textwidth]{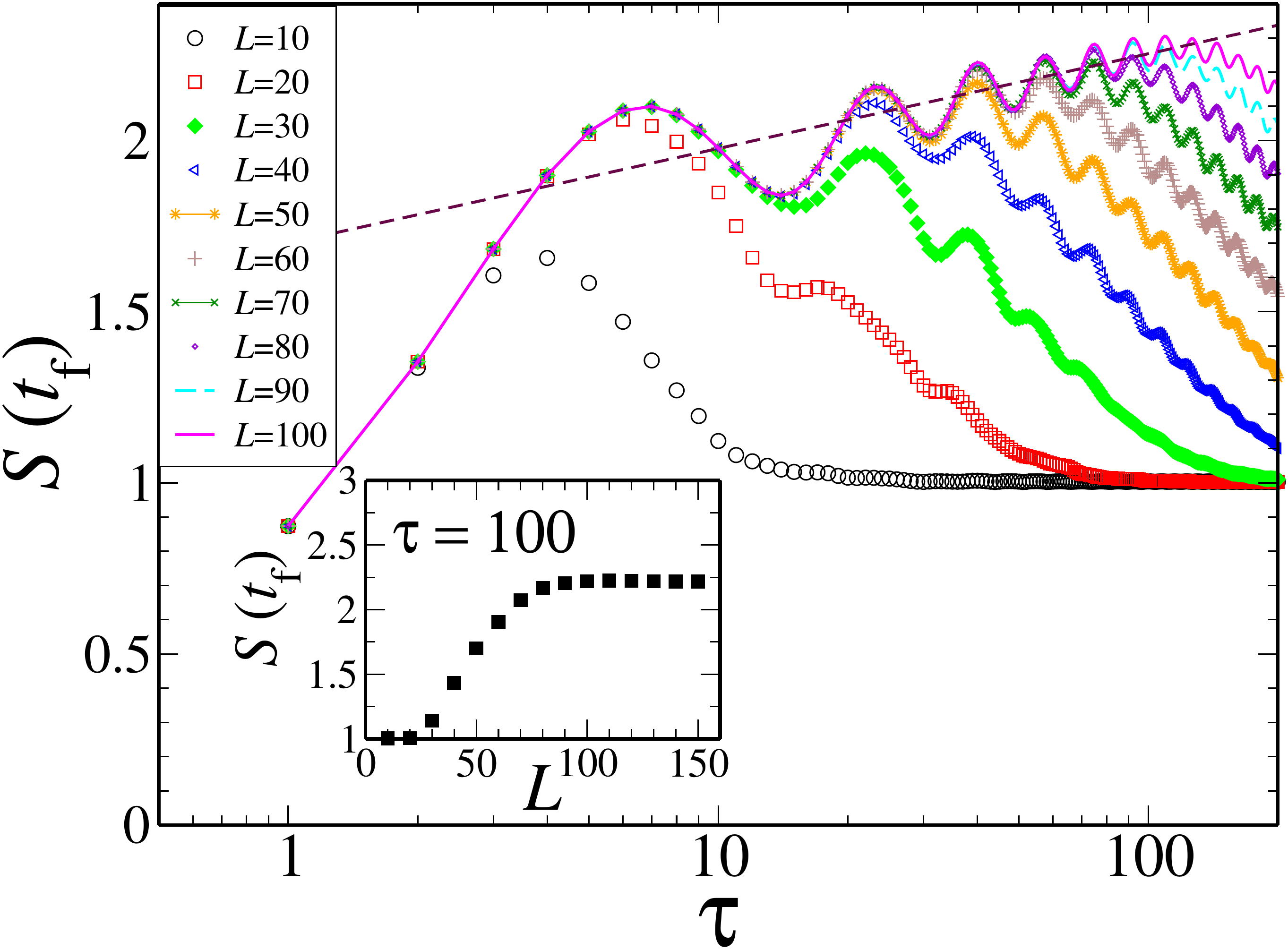}
   \caption{Main panel: entanglement entropy at the final instant of the evolution for $\tau\in[1,200]$ at different system sizes ($L=10\div100$, from bottom to top). Dashed maroon line: $y=\frac{1}{12}\log_2x+\mbox{const}$. Inset: entanglement entropy at the final instant of the evolution for $\tau=100$ as a function of the system size $L$.}\label{KZ-EE+SG}
 \end{figure}

 Any quantity that is directly related to the correlation length is suitable to a Kibble-Zurek analysis. In particular, close to a conformal critical point of conformal charge $c$,  
 the entanglement entropy has been shown by Calabese and Cardy to diverge, in an infinite system, as~\cite{CalabreseCardy2004}:
 \begin{equation}
  S=\frac{c}{6}\log_2\xi+\mbox{const.}\label{KZ-EE}
 \end{equation}
 In particular, we remark that, because of the infinite size of the system, subsystem $A$ possesses just one effective boundary. The entanglement entropy after the quench is therefore easily seen to scale as\cite{Pollmann_PRE10}
 \begin{equation}
  S=\frac{c\nu}{6(1+z\nu)}\log_2\tau+\mbox{const.}
 \end{equation}
 The prefactor of the logarithm is 1/24, since in the Ising case $\nu=z=1$ and $c=1/2$.
This clearly holds in the thermodynamic limit, where the gap is strictly closed at the critical point. In our case, at finite size, we expect some deviations from the Kibble-Zurek behaviour for large $\tau$. We plot the results we obtain in Fig.~\ref{KZ-EE+SG}:  as expected, we observe a progressive breakdown of the Kibble-Zurek prediction lowering $L$ (see also the inset of Fig.~\ref{KZ-EE+SG}). 
 A few other remarks are in order: first, Eq.~\ref{KZ-EE} has to be modified, since, because of its finite size, subsystem $A$ possesses two boundaries; therefore, Eq.~\ref{KZ-EE} is modified by doubling the prefactor of the logarithm\cite{CalabreseCardy2004} (see also Ref. \onlinecite{Cincio_PRA07}). Moreover, it is evident that the logarithmic behaviour expected from the Kibble-Zurek mechanism is superimposed to an oscillating behaviour, as already observed in Ref. \onlinecite{Pollmann_PRE10}: it is clearly a reflex of the oscillating structure of the entanglement entropy as a function of time, studied in Sec.~\ref{fast} and \ref{slow}.  Third, we observe that, for small values of $\tau$, the curves at different sizes are practically coincident. This coincidence is lost for larger values of $\tau$'s, depending on $L$: the velocities at which this coincidence is observed are the ones at which the physics is practically the one of the thermodynamic limit. For example, at $L=50$, the physics is practically 
the thermodynamic limit one up to $\tau\approx15$.

Finally, we note that, remarkably, the $\tau$'s that correspond to the passage from the fast to the slow regime (the $\tau$'s for which the crossing between the first and the fourth eigenvalue of the reduced density matrix begin to disappear), correspond to the breakdown of the Kibble-Zurek, or, equivalently, thermodynamic-limit physics. This fact could be verified by a direct thermodynamic-limit investigation (as, e.g., in Ref. \onlinecite{Pollmann_PRE10}), and could represent, in principle, a very simple tool to check the equivalence between finite-size and thermodynamic-limit physics.

 \subsubsection{Schmidt gap}

 As already mentioned above, the Schmidt gap $\Delta_S$ is defined as the difference between the two highest eigenvalues of the reduced density matrix. 
 It has been very recently shown~\cite{Dechiara_PRL12} to be related to the correlation length 
 i.e.,
 \begin{equation}
  \Delta_S\approx\xi^{-z}
 \end{equation}
 and therefore its Kibble-Zurek scaling is
 \begin{equation}
  \Delta_S\approx\tau^{-\frac{z\nu}{1+z\nu}}\label{KZ-SG}
 \end{equation}
 In our case the exponent of $\tau$ in Eq.~\ref{KZ-SG} is $-1/2$.
 \begin{figure}[t]
    \includegraphics[width=0.45\textwidth]{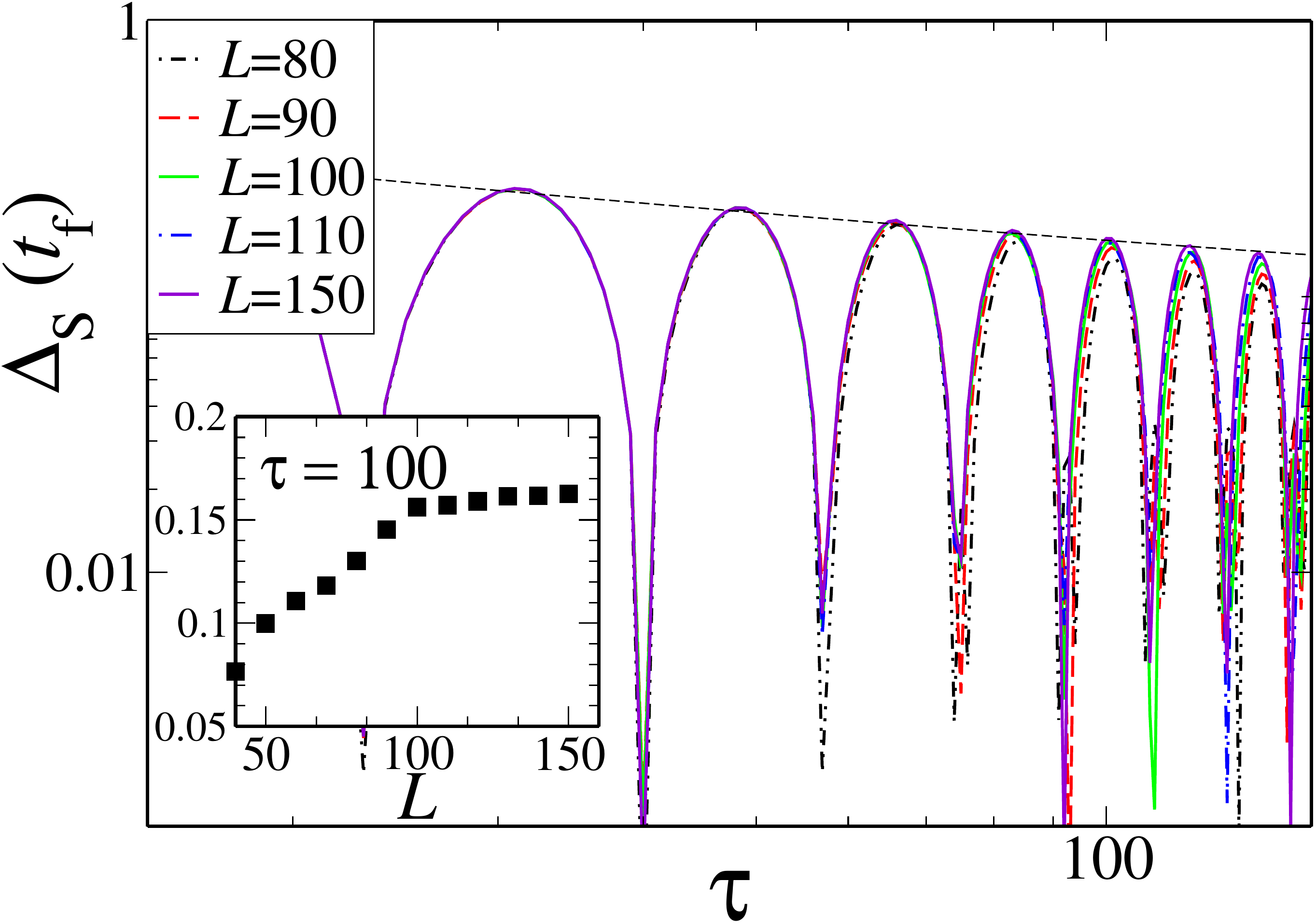}
\caption{Main panel: Schmidt gap at the final instant of the evolution for $\tau\in[15,150]$ at different system sizes ($L=80\div150$, from bottom to top). Black line: $y\approx x^{-1/2}$.  Inset: Schmidt gap at the final instant of the evolution for $\tau=100$ as a function of the system size $L$.}\label{KZ-EE+SG_2}
 \end{figure}

In Fig.~\ref{KZ-EE+SG_2} we present the data for the scaling of the Schmidt gap  at the end of the ramping as a function of $\tau$. 
At fixed $L$ the shape of each curve shows cusps as a function of $\tau$ (which correspond to crossings of the first two eigenvalues of the reduced density matrix) superimposed to an overall power-law decay. First we comment on these non-analyticies and then discuss the scaling with $\tau$. 
We suggest a possible relation of the zeroes of the Schmidt gap with the dynamical phase transitions first discussed by Heyl et al.~\cite{Heyl_PRL13}. These transitions manifest themselves as periodic non-analicities of the free-energy density in the thermodynamic limit. Although in the latter paper only the case of a sudden quench is explicitly considered, in the same reference a connection is done with the non-analytic behavior of the Loschmidt echo~\cite{Pollmann_PRE10}. In that work  they consider  the case of finite-velocity ramping followed by an evolution with constant magnetic field~\cite{Pollmann_PRE10}, i.e.  a protocol similar to Eq.~\ref{waiting}. We have verified that applying the same protocol, during the evolution at fixed magnetic field,the Schmidt gap as a function of time shows periodic cusps, with the period depending on $\tau$, in analogy with the Loschmidt echo~\cite{Pollmann_PRE10}. In the case of a sudden quench, cusps of the Schmidt gap 
are also found in Ref.~\onlinecite{TorlaiTagliacozzoDeChiara2013} by Torlai et al, who also suggest a relation with the dynamical phase transitions.
 We then suggest that the non-analyticities of the Schmidt gap at $t_{f}=\tau|h_{f}-h_{i}|$ for different $\tau$ shown in Fig.~\ref{KZ-EE+SG_2} may be interpreted as an early-time indication of dynamical phase transition. 

Concerning the power-law scaling as a function of $\tau$,
as expected from the behaviour of the entanglement entropy, for each $\tau$ the Schmidt gap tends to converge to a finite value increasing the size $L$, as shown in the inset
of Fig.~\ref{KZ-EE+SG_2}. 
We expect the scaling function of Eq.~\ref{KZ-SG} to be compatible with the numerical results.  What we actually find is something more, i.e., 
that Eq.~\ref{KZ-SG} almost perfectly interpolates the maxima of the curves at large system sizes. 

\section{Ferromagnet to paramagnet}\label{IV}

 In this Section, we perform the same analysis discussed in Sec.~\ref{pf}, but now for the transition from the ferromagnetic to the paramagnetic region. This transition is much less studied in literature than the previous one (but see Ref.~\onlinecite{Dziarmaga2005}), since it is not related to the Kibble-Zurek mechanism; 
however, it is still interesting to consider it in the present work.

\subsection{Initial structure of the entanglement spectrum}

 Let us consider now the Hamiltonian in Eq.~\ref{eq:Ising}, for simplicity, with $h=0$. The ground space takes the form $\mbox{span}\left\{\left|up\right>,\left|down\right>\right\}$, with $\left|up\right>\equiv\prod_{j=1}^L\left|\uparrow\right>_j$ and $\left|down\right>\equiv\prod_{j=1}^L\left|\downarrow\right>_j$. At finite size, the ground state always belong to the $\alpha=1$ sector (see Appendix~\ref{solution}): therefore, since $\sigma^x\left|\uparrow\right>=\left|\downarrow\right>$ and {\it viceversa}, the it is easily seen to be (up to a phase)
 \begin{equation}
  \left|GS,+\right>\equiv\frac{1}{\sqrt{2}}\left(\left|up\right>+\left|down\right>\right)
 \end{equation}

 The zero-temperature density matrix of the system is therefore
 \begin{equation}
  \rho=\left|GS,+\right>\left<GS,+\right|
 \end{equation}
 and the correspondent reduced density matrix looks, for general $A$,
 \begin{equation}
  \rho_A=\frac{1}{2}\left(\begin{array}{ll}
   \left|up\right>_A, & \left|down\right>_A
  \end{array}\right)\mathbb{I}_2\left(\begin{array}{l}
   _A\left<up\right|\\
   _A\left<down\right|
  \end{array}\right)
 \end{equation}
 The initial entanglement structure is much simpler than in the other case, just two equally weighted eigenstates playing a role.

\subsection{Dynamics of entanglement entropy and entanglement spectrum}

 We work with $L=50$ and set the initial value of the magnetic field to $h_i=0.5$. The reason of this choice is that  
 the entanglement spectrum is practically identical to the predicted one at $h=0$, with just two equal eigenvalues different from zero,
 but we can take the advantage of solving the ODE's to shorter times at fixed $\tau$. 
 The final value of the magnetic field is set to $h_f=1.5$. 

\begin{figure}[t]
 \includegraphics[width=0.49\textwidth]{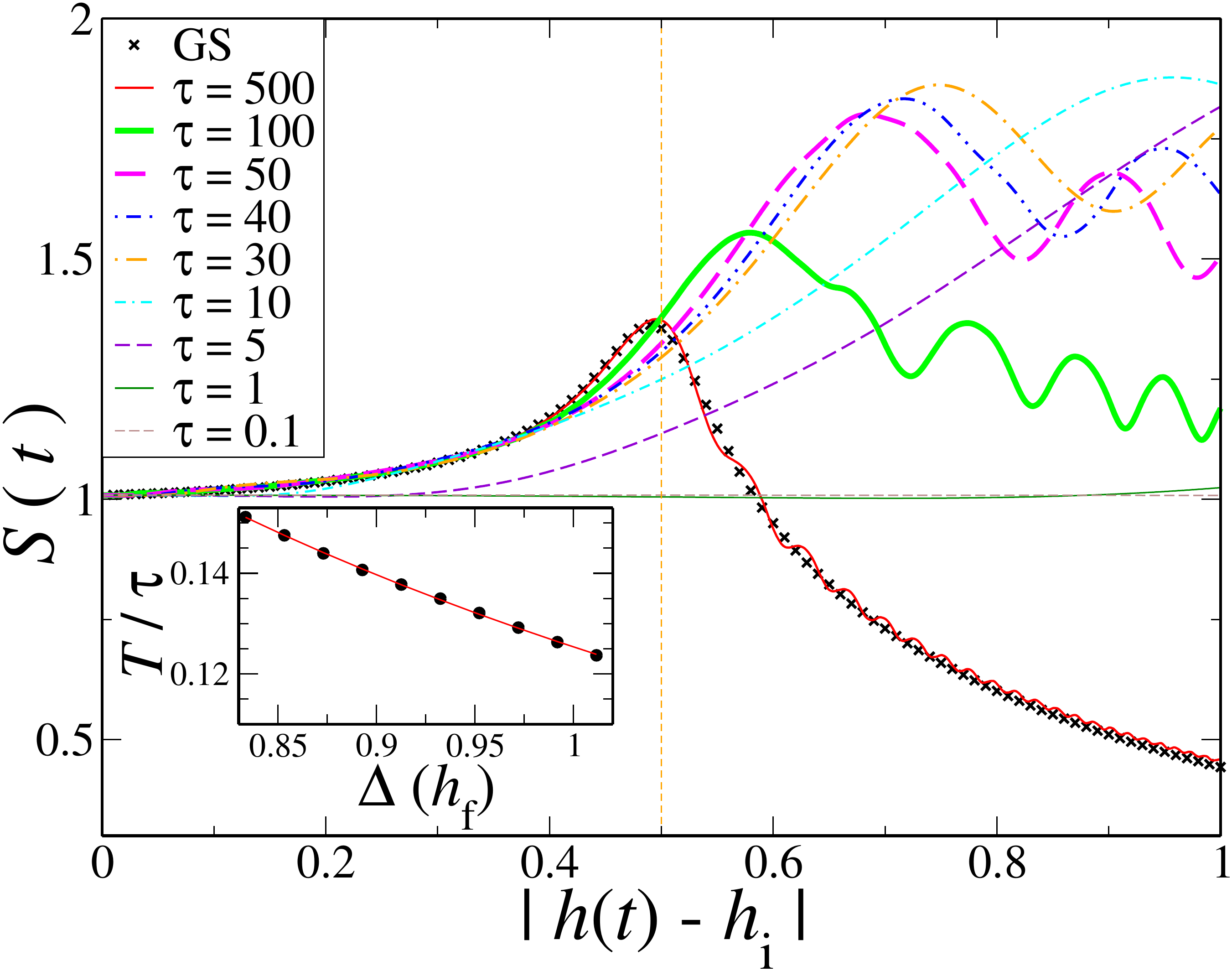}
 \caption{Main panel: dynamics of the entanglement entropy $S(t)$ for $L=50$, $h_i=0.5$, $h_f=1.5$ and for different values of $\tau$ (dashed vertical line: location of the critical point). Inset, black dots: period of oscillation of the entropy as a function of the gap at $h=h_{f}$ ($h_{i}=0.5$, $h_{f}=1.41, 1.42,\dots1.5$) with $\tau=50$. Red line: fit with $y=a_{0}+a_{1}/x$. }\label{fig:entfp}
\end{figure}

\begin{figure*}[t]
%
 \includegraphics[width=0.8\textwidth]{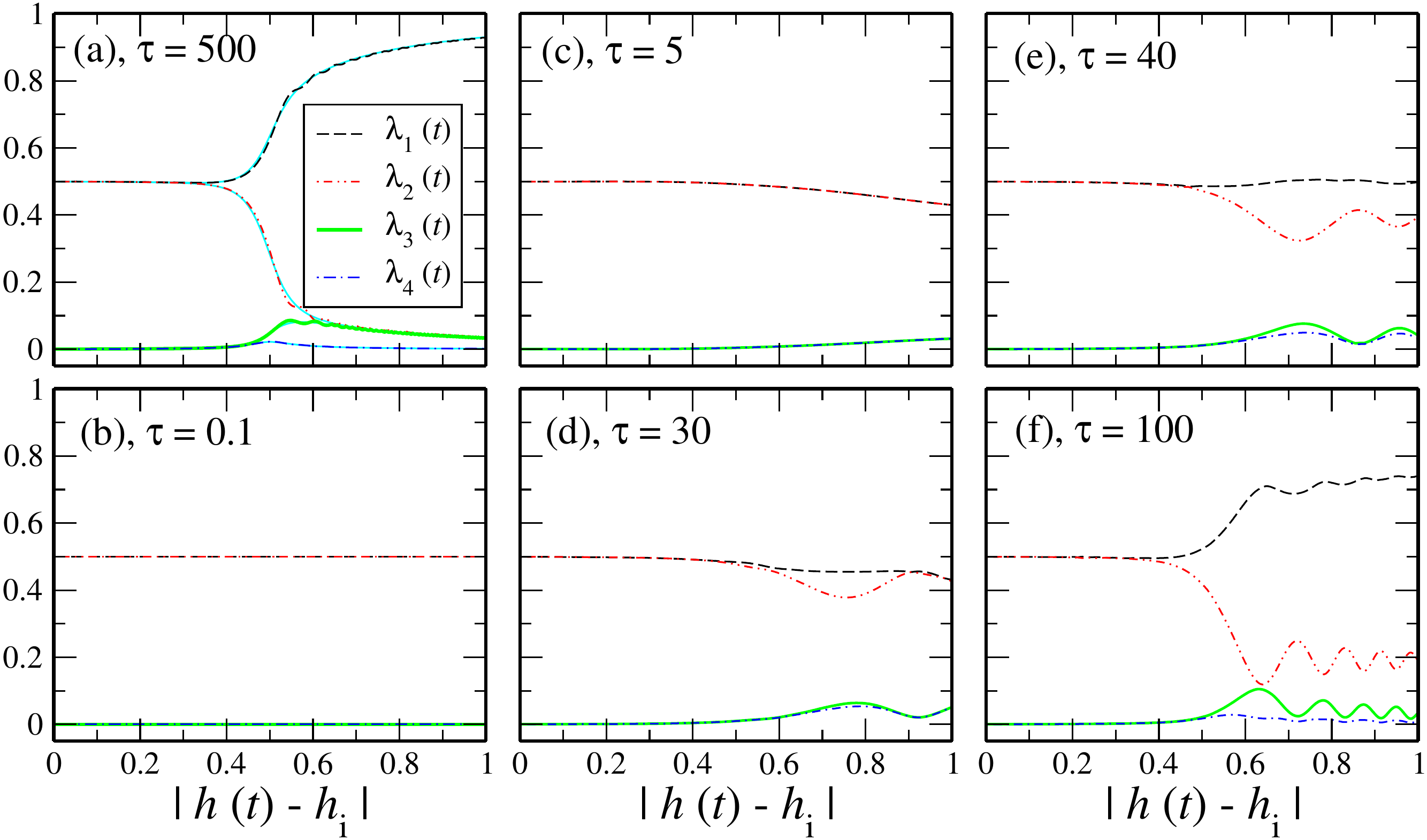}
 \caption{Dynamics of the entanglement spectrum with $L=50$, $h_i=0.5$ and $h_f=1.5$. Red/blue/green/brown line: dynamical first/second/third/fourth eigenvalue of the reduced density matrix of the half chain. Panels (a)-(f): $\tau=500$, $0.1$, $5$, $30$, $40$, $100$. The cyan lines in panel (a) show the 
 ground-state values of the first four eigenvalues. In panels (b) and (c) black and red, green and blue data are almost coincident.}\label{f->p_fig}
\end{figure*}

 Similarly to what found in Sec.~\ref{pf},
 we observe the two limiting regimes, the adiabatic and the sudden one, 
 respectively for large and small values of $\tau$ (see Figs. \ref{fig:entfp} and \ref{f->p_fig}) and,  at intermediate values of $\tau$ we find the regimes described in Sec. \ref{pf}. Remarkably, as already mentioned, oscillations in the entanglement entropy are present (see Fig.~\ref{fig:entfp}) and, similarly to  what we discussed in Sec.~\ref{pf}, their origin can be traced to the partially excited nature of the wave function after passing the critical point. Indeed, we can let the system evolve with protocol of Eq.~\ref{waiting} and study the period of the oscillations of the entropy when $t>(h_{f}-h_{i})\tau$ as a function of the dynamical gap at the final value of the magnetic field $h=h_{f}$. 
 Analogously to the paramagnetic to ferromagnetic quench, we find that the period scales inversly proportional to the gap, as we show in the inset of Fig.~\ref{fig:entfp}.

\begin{figure}[t]
 \includegraphics[width=0.49\textwidth]{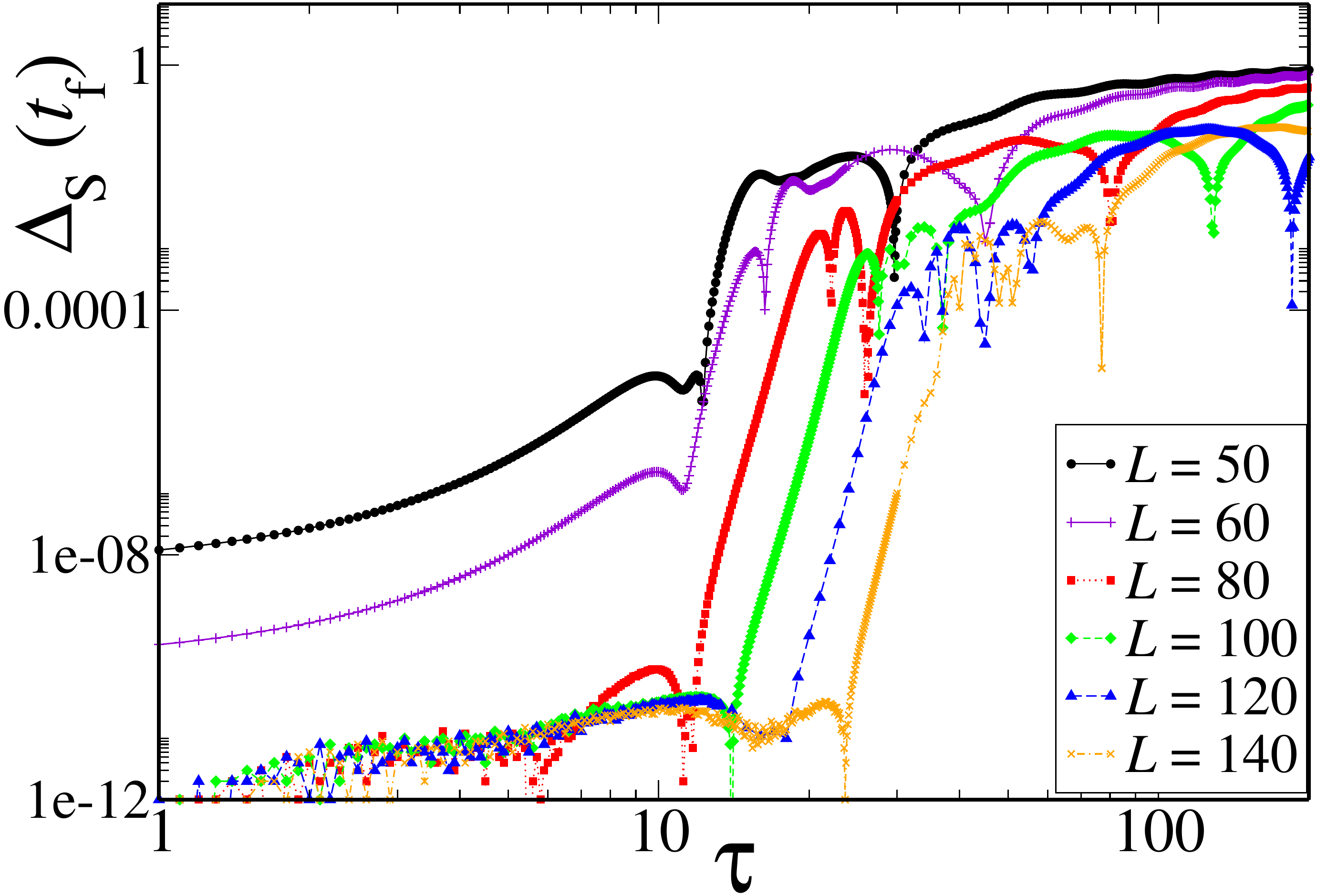}
 \caption{Schmidt gap at the final instant of the evolution for $\tau\in[1,200]$ at different system sizes for a ramping from $h_{i}=1.5$ to
 $h_{f}=0.5$.}\label{fig:schfp}
\end{figure}

The four largest eigenvalues of the entanglement spectrum are degenerate in pairs before crossing the critical point (see all the panels of Fig.~\ref{f->p_fig} for $|h(t)-h_{i}|<0.5$). 
After crossing the transition, the structure of the eigenvalues changes, so that only the second and the third eigenvalues are degenerate in the adiabatic limit, as we show in Fig. \ref{f->p_fig}(a).  For slow rampings, these eigenvalues cross each other with a regular pattern analogous to that of Fig.~\ref{p->f_fig}(a), (f). The Schmidt gap is  finite in this limit, apart from cusps originating from the oscillation of $\lambda_{1}$ and $\lambda_{2}$, as it is shown in Fig.~\ref{fig:schfp} for $\tau\gtrsim 30$. 
 On the contrary, if the ramping of the magnetic field is fast, the two pairs of eigenvalues remain almost degenerate,
so that the Schmidt gap is several orders of magnitude smaller than in the adiabatic limit, as shown in Fig.~\ref{fig:schfp} for $\tau\lesssim10$.    

We conclude this section with two comments: first, going from the sudden to the adiabatic limit, the Schmidt gap increases  sharply (see Fig.~\ref{fig:schfp} 
for $\tau \sim 10\div 40)$, second, comparing Figs. ~\ref{f->p_fig} and ~\ref{fig:schfp} at $L=50$ , $\tau=30$, it emerges that the crossing of the first and the second eigenvalue, giving rise to the cusp in the Schmidt gap, is isolated, because it corresponds to a narrow region of $\tau$ where the degeneracy of $\lambda_{1}$ and $\lambda_{2}$ is only slightly lifted, but for slower quenches these crossings disappear. 

 %
\section{Conclusions}\label{conc}
In this work we have examined the dynamical evolution of the quantum Ising chain in a transverse magnetic field by looking at entanglement entropy and entanglement spectrum, in particular in the case of a ramping from the paramagnetic to the ferromagnetic phase, and {\it viceversa}. We made the Hamiltonian time-dependent by letting the magnetic field to vary linearly in time with a varying time scale $\tau$,
obtaining three qualitatively different regimes: an adiabatic one (large $\tau$) when the system evolves according the instantaneous ground state, a sudden quench (small $\tau$) when the system is essentially frozen to its initial state and an intermediate one, where complicated behaviours occur. In particular, the most interesting feature we observe is the arising, in this regime, of dynamical multiple crossings of the first Schmidt eigenvalues: this effect is partially understood by means of an analysis of the fine structure of the corresponding eigenvectors, even if its general explanation, together with its observation in different models, is still missing. However, the physics of the dynamical evolution is well understood by looking at the behaviour in time of the entanglement spectrum, starting from which one can study both universal quantities (scaling exponents) and physical phenomena, such as the Kibble-Zurek mechanism, that may manifest during the evolution.

We may conclude that entanglement entropy and entanglement spectrum seem to be, for the dynamical evolution as in the static case, a powerful tool to investigate the physics of a closed quantum many body system crossing a phase transition at $T=0$. We have explicitly used this technique to study a paradigmatic exactly integrable system such as the quantum Ising chain in a transverse magnetic field, but investigation is under way to examine different situations where we either break integrability and/or introduce disorder.

After this work was submitted, two independent studies of the dynamics of the entanglement spectrum in one-dimensional models appeared~\cite{TorlaiTagliacozzoDeChiara2013,ZRLLT2014}.

\acknowledgments
 We thank D. Bianchini, T. Caneva, C. Degli Esposti Boschi, M. Dalmonte, L. Ferrari, A. Lazarides, F. Ortolani, G. Zonzo and the participants to the mini-workshop ``Collective quantum phenomena: methods and tools from entanglement theory'', organized by F. Illuminati in Salerno, for enlightening discussions. E.C. acknowledges  financial support from DPG through project SFB/TRR21 and Dipartimento di Fisica e Astronomia dell'Universit\`a di Bologna for hospitality.
\appendix
\section{Exact solution of the Ising model}\label{solution}
In this Appendix we show how to diagonalize the Hamiltonian in Eq.~\ref{eq:Ising}. We follow Ref.~\onlinecite{Franchini2011} quite closely.

By defining the raising and lowering operators $ \sigma_j^\pm\equiv(\sigma_j^z\mp i\sigma_j^y)/2$,
Eq.~\ref{eq:Ising} reads:
\beq\label{eq:ladderH}
 H=-\frac{1}{2}\sum_{j=1}^L\left[\left(\sigma_j^+\sigma_{j+1}^++\sigma_j^+\sigma_{j+1}^-+{\rm h.c.}\right)+2h\sigma_j^+\sigma_j^-\right]+\frac{Lh}{2}
\eeq
Performing a Jordan-Wigner transformation by means of 
\beq
\begin{array}{l}
  c_j\equiv\prod_{k=1}^{j-1}\left(2\sigma_k^+\sigma_k^--1\right)\sigma_j^+\\
  \\
  c_j^\dagger\equiv\sigma_j^-\prod_{k=1}^{j-1}\left(2\sigma_k^+\sigma_k^--1\right)
 \end{array}
\eeq
allows to rewrite the Hamiltonian Eq.~\ref{eq:ladderH} in fermionic form:
\beq
\begin{split}
  H&=-\frac{1}{2}\sum_{j=1}^{L-1}\left[c_{j+1}^\dagger c_j+c_{j+1}c_j+{\rm h.c.}\right]+\\
  &+\frac{\alpha}{2}\left[c_1^\dagger c_L+c_1c_L + {\rm h.c.}\right]+h\sum_{j=1}^Lc_j^\dagger c_j-\frac{Lh}{2}
 \end{split}
\eeq
where $\alpha\equiv\prod_{j=1}^L(1-2c_j^\dagger c_j)=\prod_{j=1}^L\sigma_j^x$. 
It is easy to show that $\alpha$ commutes with $H$, and therefore it is a constant of motion; moreover, $\alpha^2=1$, so that $\alpha=\pm1$. 
As it is manifest from its definition, the case $\alpha=\pm1$ corresponds to the case in which in the chain has an even/odd number of down spins is present, and, in fermionic language, to a chain with antiperiodic/periodic boundary conditions (APBC/PBC) and an even/odd number of fermions. We choose to work in the sector of even parity in the number fermions, i.e. $\alpha=1$, being, at finite size, the ground state of the model always in this sector. One ends up with
\begin{equation}
 H=-\frac{1}{2}\sum_{j=1}^L\left[\left(c_{j+1}^\dagger c_j+c_{j+1}c_j+{\rm h.c.}\right)-2hc_j^\dagger c_j\right]-\frac{Lh}{2}
\end{equation}
with fermions satisfying APBC.
The diagonalization proceeds by means of a Fourier transform
\begin{equation}
 c_j\equiv\frac{e^{i\pi/4}}{\sqrt{L}}\sum_{m=0}^{L-1}e^{ip_mj}d_m\;,
\end{equation}
with $p_m\equiv 2\pi(m+1/2)/L$, in order to automatically implement the APBC. 
With some algebra, it is possible to show that the Hamiltonian takes the form
\begin{equation}
 H=\frac{1}{2}\sum_{m=0}^{L-1}\left(d_m^\dagger,d_{L-m-1}\right)M_m\left(\begin{array}{c}
  d_m \\ d^\dagger_{L-m-1}
 \end{array}\right)
\end{equation}
with
\begin{equation}
 M_m\equiv\left(\begin{array}{cc}
  A_m & -B_m \\
  -B_m & -A_m
 \end{array}\right)
\end{equation}
and
\begin{equation}
 A_m\equiv h-\cos p_m,\ \ B_m\equiv\sin p_m
\end{equation}
that, remarkably, satisfy $A_{L-m-1}=A_m$, $B_{L-m-1}=-B_m$, i.e., the Hamiltonian decouples into the sum of $L$ non-interacting modes, each one independently diagonalizable.\\
The last step of the procedure consists of a Bogolyubov transformation, which puts each $M_m$ in diagonal form. 
The eigenvalues of each $M_m$ are given by the two values $\pm E_m$, with
\begin{equation}
 E_m=\sqrt{A_m^2+B_m^2}
\end{equation}
and the orthogonal transformation $U_m$ making $M_m$ diagonal, i.e., giving $U_m^\dagger M_mU_m=\mbox{diag}(E_m,-E_m)$, is given by
\begin{equation}
 U_m\equiv\left(\begin{array}{cc}
  u_m & v_m \\
  -v_m & u_m
 \end{array}\right)
\end{equation}
where
\begin{equation}
 u_m=\frac{-(-1)^m\frac{A_m+E_m}{B_m}}{\sqrt{1+\left(\frac{A_m+E_m}{B_m}\right)^2}},\ \ v_m=\frac{-(-1)^m}{\sqrt{1+\left(\frac{A_m+E_m}{B_m}\right)^2}}
\end{equation}
satisfying $u_{L-m-1}=u_m$, $v_{L-m-1}=-v_m$. The diagonalizing operators are
\begin{equation}\label{bog}
 \left(\begin{array}{c}
  b_m \\ b_{L-m-1}^\dagger
 \end{array}\right)\equiv U_m\left(\begin{array}{c}
  d_m \\ d_{L-m-1}^\dagger
 \end{array}\right)
\end{equation}
and the orthogonality of $U_m$ ensures their fermionic nature. The Hamiltonian takes, by means of the inverse of Eq.~\ref{bog}, the final form
\begin{equation}
 H=\sum_{m=0}^{L-1}E_m\left(b_m^\dagger b_m-\frac{1}{2}\right)
\end{equation}
and its ground state is, for $\alpha=1$, the vacuum state $\left|GS\right>$ such that $b_m\left|GS\right>=0$. Excited states, in the APBC sector, are obtained by applying couples of Bogolyubov creation operators on $\left|GS\right>$.

\section{Dynamics in the Ising model}\label{dynamics}
In this Appendix, we show how to describe the dynamics of a state according to the Hamiltonian in Eq.~\ref{eq:Ising}. We follow the procedure of Ref.~\onlinecite{CanevaFazioSantoro2007}.

The time evolution of the system in Eq.~\ref{eq:Ising} is described by the Heisenberg equation for the $c$ operators:
\begin{equation}
 i\frac{d}{dt}c_{j,H}(t)=\left[c_{j,H}(t),H_{j,H}(t)\right]
\end{equation}
which can be rewritten as:
\begin{equation}
 i\frac{d}{dt}c_{j,H}(t)=\sum_{k=1}^L\left[A_{jk}(t)c_{k,H}(t)+B_{jk}(t)c^\dagger_{k,H}(t)\right]
\end{equation}
with
\begin{equation}
 \begin{split}
  A_{jk}(t) &\equiv h(t)\delta_{jk}-\frac{1}{2}\left(\delta_{j,k+1}+\delta_{j+1,k}-\delta_{j1}\delta_{kL}-\delta_{jL}\delta_{k1}\right)\\
  B_{jk}(t) &\equiv-\frac{1}{2}\left(\delta_{j+1,k}-\delta_{j,k+1}+\delta_{j1}\delta_{kL}-\delta_{jL}\delta_{k1}\right)
 \end{split}
\end{equation}
In order to solve such an equation, we make the following ansatz, known as {\it time-dependent Bogolyubov transformation}:
\begin{equation}
 c_{j,H}(t)\equiv\sum_{m=0}^{L-1}\left[u_{jm}(t)b_m+v_{jm}^*(t)b_m^\dagger\right]\label{t-bog}
\end{equation}
with the initial conditions $u_{jm}(0)=u_{jm}$ and $v_{jm}(0)=v_{jm}$ given by the exact solution:
\begin{equation}
 \begin{split}
  u_{jm} &\equiv\frac{1}{\sqrt{L}}e^{i\left(p_mj+\frac{\pi}{4}\right)}u_m,\\
  v_{jm} &\equiv\frac{1}{\sqrt{L}}e^{i\left(p_mj+\frac{\pi}{4}\right)}v_m
 \end{split}
\end{equation}
By putting the ansatz of Eq.~\ref{t-bog} in the Heisenberg equation, we come to the set of linear coupled ODE's
\begin{equation}\label{ODE}
 \begin{split}
  i\frac{d}{dt}u_{jm}(t)&=\sum_{k=1}^L\left[A_{jk}(t)u_{km}(t)+B_{jk}(t)v_{km}(t)\right]\\
  -i\frac{d}{dt}v_{jm}(t)&=\sum_{k=1}^L\left[B_{jk}(t)u_{km}(t)+A_{jk}(t)v_{km}(t)\right]
 \end{split}
\end{equation}

\section{Bipartite quantities in free fermionic systems}\label{bip}

In this Appendix we review to compute the entanglement entropy and the entanglement spectrum for free fermionic system.

As it is known from recent 
literature~\cite{VLRK2003,Peschel_JPA03} (see also Refs.~\onlinecite{Berganza_JSTAT12,Taddia_Phd}), for fermionic biquadratic (static) Hamiltonians the density matrix
can be obtained from correlation functions. In order to evaluate the time evolution of the entanglement entropy and spectrum we need a step forward, which is the introduction 
of Majorana fermions:
\begin{align}
\bar c_{2m-1}=&c^{\dagger}_{m}+c_{m}\\
\bar c_{2m}=&i(c^{\dagger}_{m}-c_{m})
\end{align}
which satisfy anticommutation rules $\{\bar c_{r},\bar c_{s}\}=2\delta_{rs}$.
The correlation matrix of the Majorana fermions has the form:
\beq
\langle\bar c_{r}\bar c_{s}\rangle=\delta_{r,s}+i\Gamma_{rs}
\eeq
where $r,s=1,\cdots,2\ell$. The matrix $\Gamma_{rs}$ is antisymmetric and its eigenvalues are purely
imaginary $\pm i \nu_{r}$, $r=1,\ell$. It can be shown that this matrix describes 
a set of uncorrelated  (true) fermions $\{ a_{m}\}$ satisfying:
\beq
\langle a_{m}a_{n}\rangle=0,\qquad \langle a^{\dagger}_{m}a_{n}\rangle=\delta_{mn}\frac{1+\nu_{n}}{2}\;.
\eeq
Each of the $\ell$ blocks is then in the state $\rho_{j}=p_{j}a^{\dagger}_{j}\ket{0}\bra{0}a_{j}+(1-p_{j})\ket{0}\bra{0}$, 
with $p_{j}=(1+\nu_{j})/2$
so that the entropy is the sum of the single-particle entropies, thus yielding for the reduced $\ell$-site system:
\beq\label{eq:entropy}
S(\ell) = \sum_{j=1}^{l}H_{2}\left(\frac{1+\nu_{j}}{2}\right)\;,
\eeq
where $H_{2}(x)\equiv-x \log_2 x -(1-x)\log_2 (1-x)$.
The eigenvalues $\lambda_{j}\;,\quad j=1,\cdots,2^{\ell}$ of the reduced density matrix can in principle be found by taking properly chosen products of either $p_{j}$ or $\left(1-p_{j}\right)$,
with $j=1,\cdots,\ell$~\cite{CalabreseLefevre2008}. 

The procedure described above works equally well for the time-dependent case, provided that the 
Majorana fermions are constructed using the time-evolved true fermions $c_{i,H}(t)$. In this way 
we can obtain the time-dependent entropy $S(\ell,t)$ and entanglement spectrum  $\lambda_{i}(\ell,t)$.




\begin{thebibliography}{99}
\bibitem{PSSV2011} A. Polkovnikov, K. Sengupta, A. Silva, M. Vengalattore, Rev. Mod. Phys. {\bf 83}, 863 (2011).
\bibitem{LiebSchultzMattis1961} E. Lieb, T. Schultz, D. Mattis, Ann. Phys. {\bf 16}, 407 (1961).
\bibitem{Sachdev:book} S. Sachdev, {\it Quantum Phase Transitions}, 2nd edition, Cambridge University Press, Cambridge (2011).
\bibitem{Franchini2011} F. Franchini, {\it Notes on Bethe Ansatz Techniques}, http://people.sissa.it/\verb1~1ffranchi/BAClass.html/ (2011).
\bibitem{DKSV2004} A. J. Daley, C. Kollath, U. Schollwoeck, G. Vidal, J. Stat. Mech.: Theor. Exp. 2004, P04005.
\bibitem{WhiteFeiguin2004} S. R. White, A. E. Feiguin, Phys. Rev. Lett. {\bf 93}, 076401 (2004).
\bibitem{FeiguinWhite2005} A. E. Feiguin, S. R. White, Phys. Rev. B {\bf 72}, 020404(R) (2005).
\bibitem{Vidal2004} G. Vidal, Phys. Rev. Lett. {\bf 93}, 040502 (2004).
\bibitem{CiracVerstraete2009} J. I. Cirac, F. Verstraete, J. Phys. A: Math. Theor. {\bf 42}, 504004 (2009).
\bibitem{AFOV2008} L. Amico, R. Fazio, A. Osterloh, V. Vedral, Rev. Mod. Phys. {\bf 80}, 517 (2008).
\bibitem{XavierAlcaraz2011} J. C. Xavier, F. C. Alcaraz, Phys. Rev. B {\bf 84}, 094410 (2011).
\bibitem{Nishimoto2011} S. Nishimoto, Phys. Rev. B {\bf 84}, 195108 (2011).
\bibitem{DalmonteErcolessiTaddia2011} M. Dalmonte, E. Ercolessi, L. Taddia,  Phys. Rev. B {\bf 84}, 085110 (2011).
\bibitem{DalmonteErcolessiTaddia2012} M. Dalmonte, E. Ercolessi, L. Taddia,  Phys. Rev. B {\bf 85}, 165112 (2012).
\bibitem{Ercolessietal} E. Ercolessi, S. Evangelisti, F. Franchini, F. Ravanini, Phys. Rev. B {\bf 83}, 012402 (2011).
\bibitem{Citroetal} X. Deng, R. Citro, E. Orignac, A. Minguzzi, L. Santos, New J. Phys. {\bf  15}, 045023 (2013).
\bibitem{CanevaFazioSantoro2008} T. Caneva, R. Fazio, G. E. Santoro, Phys. Rev. B {\bf 78}, 104426 (2008).
\bibitem{CalabreseCardy2005} P. Calabrese, J. Cardy, J. Stat. Mech.: Theor. Exp. 2005, P04010.
\bibitem{Pollmann_PRE10} F.~Pollmann, S.~Mukerjee, A.~G.~Green, and J.~E.~Moore, Phys. Rev. E {\bf 81}, 020101(R) (2010).
\bibitem{Cincio_PRA07} L.~Cincio, J.~Dziarmaga, M.~M.~Rams, and W.~H.~Zurek, Phys. Rev. A {\bf 75}, 052321 (2007).
\bibitem{Kibble} T. W. B. Kibble, J. Phys. A {\bf 9}, 1387 (1976); Phys. Rep. {\bf 67}, 183 (1980).
\bibitem{Zurek} W. H. Zurek, Nature (London) {\bf 317}, 505 (1985); Acta Phys. Pol. B {\bf 24}, 1301 (1993); Phys. Rep. {\bf 276}, 177 (1996).
\bibitem{ZurekDornerZoller2005} W. H. Zurek, U. Dorner, P. Zoller, Phys. Rev. Lett. {\bf 95}, 105701 (2005).
\bibitem{Dechiara_PRL12} G.~De Chiara, L.~Lepori, M.~Lewenstein, and A.~Sanpera, Phys. Rev. Lett. {\bf 109}, 237208 (2012); 
L.~Lepori, G.~De Chiara, and A.~Sanpera, Phys. Rev. B {\bf 87}, 235107 (2013).
\bibitem{GMDADSI2013} S. M. Giampaolo, S. Montangero, F. Dell'Anno, S. De Siena, F. Illuminati, Phys. Rev. B {\bf 88}, 125142 (2013).
\bibitem{NielsenChuang2000} M. A. Nielsen, I. L. Chuang, \emph{Quantum Computation and Quantum Information}, Cambridge University Press (2000).

\bibitem{CalabreseEsslerFagotti2012} P. Calabrese, F. H. L. Essler, M. Fagotti, J. Stat. Mech. (2012) P07016.

\bibitem{DiFrancescoMathieuSenechal1997} P. Di Francesco, P. Mathieu, D. S\'en\'echal, \emph{Conformal Field Theory}, Springer (1997).
\bibitem{MorandiNapoliErcolessi2001} G. Morandi, F. Napoli, E. Ercolessi, {\it Statistical Mechanics: An Intermediate Course}, World Scientific (2001).
\bibitem{McCoy_PRA71}B.~M.~McCoy, E.~Barouch and D.B.~Abraham, Phys. Rev. A {\bf 4}, 2331 (1971).
\bibitem{PerkAuYang2009} J. H. H. Perk, H. Au-Yang, J. Stat. Phys. {\bf 135}, 599 (2009).
\bibitem{Messiah1999} A. Messiah, {\it Quantum Mechanics}, Dover Publications (1999).

\bibitem{lanczos} C.~Moler and C.V. Loan, SIAM Review {\bf 45}, 3 (2003); M. Hochbruck and C. Lubich, BIT {\bf 39}, 620 (1999).
\bibitem{ZRLLT2014} A. Zamora, J. Rodriguez-Laguna, M. Lewenstein, L. Tagliacozzo, arXiv:1401.7916.
\bibitem{Dziarmaga2005} J. Dziarmaga, Phys. Rev. Lett. {\bf 95} 245701 (2005).
\bibitem{Polkovnikov2005} A. Polkovnikov, Phys. Rev. B {\bf 72}, 161201(R) (2005).
\bibitem{MDDS2007} V. Mukherjee, U. Divakaran, A. Dutta, D. Sen, Phys. Rev. B {\bf 76}, 174303 (2007).
\bibitem{DamskiZurek} B. Damski, Phys. Rev. Lett. {\bf 95}, 035701 (2005); B. Damski, W. H. Zurek, Phys. Rev. A {\bf 73}, 063405 (2006).

\bibitem{CalabreseCardy2004} P. Calabrese, J. Cardy, J. Stat. Mech.: Theor. Exp. 2004, P06002.
\bibitem{Heyl_PRL13} M.~Heyl, A.~Polkovnikov, and S.~ Kehrein, Phys. Rev. Lett. {\bf 110} 135704 (2013).
\bibitem{TorlaiTagliacozzoDeChiara2013} G. Torlai, L. Tagliacozzo, G. De Chiara, arXiv:1311.5509.
\bibitem{CanevaFazioSantoro2007} T. Caneva, R. Fazio, G. E. Santoro, Phys. Rev. B {\bf 76}, 144427 (2007).
\bibitem{Peschel_JPA03} I.~Peschel, J. Phys. A: Math. Gen. {\bf 36}, L205 (2003).
\bibitem{VLRK2003} G. Vidal, J. I. Latorre, E. Rico, A. Kitaev, Phys. Rev. Lett. {\bf 90}, 227902 (2003).
\bibitem{Berganza_JSTAT12} M.~I.~Berganza, F.~C. Alcaraz, and G.~Sierra, J. Stat. Mech.: Theor. Exp. P01016 (2012).
\bibitem{Taddia_Phd} L.~Taddia, {\it Entanglement Entropies in One-Dimensional Systems}, Lambert Academic Publishing (2013); arXiv:1309.4003.
\bibitem{CalabreseLefevre2008} P. Calabrese, A. Lefevre, Phys. Rev. A {\bf 78}, 032329 (2008).






\end{thebibliography}
\end{document}